\documentclass[sigconf]{acmart}
\renewcommand\footnotetextcopyrightpermission[1]{} 
\newgeometry{left=0.625in,right=0.625in,top=0.75in,bottom=1in} 
\usepackage{mathtools}
\usepackage{breakurl}
\usepackage{makecell}
\usepackage{algorithm}
\usepackage{algpseudocode}
\usepackage{booktabs} 
\usepackage{enumitem}
\usepackage{letltxmacro}
\usepackage{multirow,array}
\usepackage{subfig}
\usepackage{scalerel}
\usepackage{listings}
\usepackage{balance}

\definecolor{gre}{RGB}{15, 140, 0}
\definecolor{evalOrange}{RGB}{239, 155, 0}

\definecolor{page1color}{RGB}{34, 185, 4}
\definecolor{page2color}{RGB}{128, 255, 104}
\definecolor{page3color}{RGB}{230, 230, 0}
\definecolor{page4color}{RGB}{109, 109, 109}
\definecolor{page5color}{RGB}{251, 0, 6}

\newcommand{\maxColor}[1]{ \textcolor[RGB]{15, 140, 0}{#1} }
\newcommand{\minColor}[1]{ \textcolor[RGB]{251, 0, 6}{#1} }
\newcommand{\maskColor}[1]{ \textcolor[RGB]{255, 255, 255}{#1} }

\begin{document}

\title{Scraping SERPs for Archival Seeds: It Matters When You Start}

\author{Alexander C. Nwala}
\affiliation{%
  \institution{Old Dominion University}
  \city{Norfolk} 
  \state{Virginia} 
  \postcode{23529}
  \country{USA}
}
\email{anwala@cs.odu.edu}

\author{Michele C. Weigle}
\affiliation{%
  \institution{Old Dominion University}
  \city{Norfolk} 
  \state{Virginia} 
  \postcode{23529}
  \country{USA}
}
\email{mweigle@cs.odu.edu}

\author{Michael L. Nelson}
\affiliation{%
  \institution{Old Dominion University}
  \city{Norfolk} 
  \state{Virginia} 
  \postcode{23529}
  \country{USA}
}
\email{mln@cs.odu.edu}
\pagestyle{empty}
\begin{abstract}
Event-based collections are often started with a web search, but the search results you find on Day 1 may not be the same as those you find on Day 7. In this paper\footnote{This is an extended version of the ACM/IEEE Joint Conference on Digital Libraries (JCDL 2018) full paper: https://doi.org/10.1145/3197026.3197056. Some of the figure numbers have changed.}, we consider collections that originate from extracting URIs (Uniform Resource Identifiers) from Search Engine Result Pages (SERPs). Specifically, we seek to provide insight about the retrievability of URIs of news stories found on Google, and to answer two main questions: first, can one ``refind'' the same URI of a news story (for the same query) from Google after a given time? Second, what is the probability of finding a story on Google over a given period of time? To answer these questions, we issued seven queries to Google every day for over seven months (2017-05-25 to 2018-01-12) and collected links from the first five SERPs to generate seven collections for each query. The queries represent public interest stories: ``healthcare bill,'' ``manchester bombing,'' ``london terrorism,'' ``trump russia,'' ``travel ban,'' ``hurricane harvey,'' and ``hurricane irma.'' We tracked each URI in all collections over time to estimate the discoverability of URIs from the first five SERPs. Our results showed that the daily average rate at which stories were replaced on the default Google SERP ranged from 0.21 -- 0.54, and a weekly rate of 0.39 -- 0.79, suggesting the fast replacement of older stories by newer stories. The probability of finding the same URI of a news story after one day from the initial appearance on the SERP ranged from 0.34 -- 0.44. After a week, the probability of finding the same news stories diminishes rapidly to 0.01 -- 0.11. In addition to the reporting of these probabilities, we also provide two predictive models for estimating the probability of finding the URI of an arbitrary news story on SERPs as a function of time. The web archiving community considers link rot and content drift important reasons for collection building. Similarly, our findings suggest that due to the difficulty in retrieving the URIs of news stories from Google, collection building that originates from search engines should begin as soon as possible in order to capture the first stages of events, and should persist in order to capture the evolution of the events, because it becomes more difficult to find the same news stories with the same queries on Google, as time progresses.
\end{abstract}
%
%

\maketitle
\section{Introduction and Background}
\sloppy
\begin{figure*}[h]%
    \centering
    \subfloat[The Google All (renamed to \textit{General}) SERP.]{{ \includegraphics[width=0.47\textwidth]{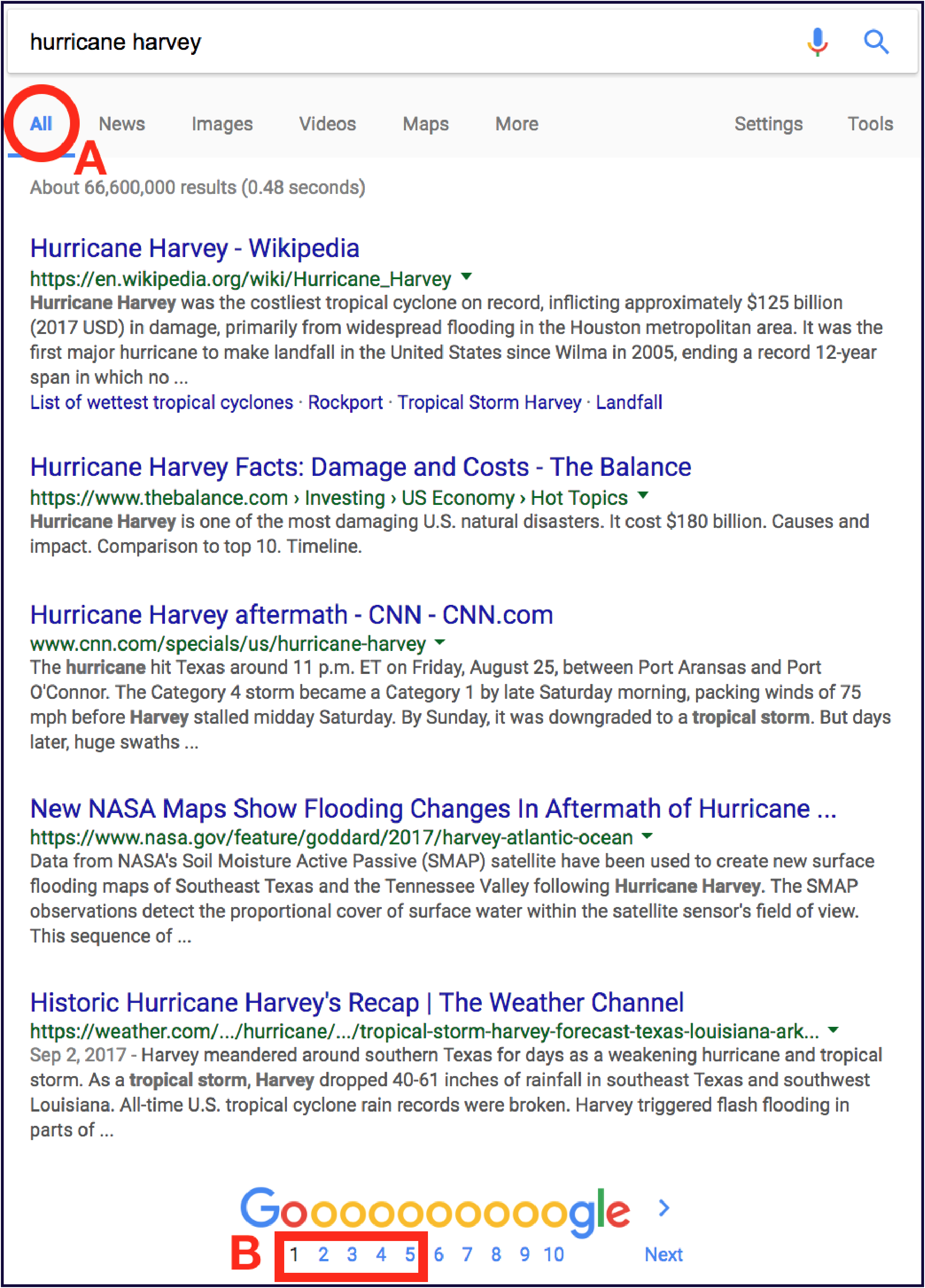} }}%
    \qquad
    \subfloat[The Google \textit{News} vertical SERP.]{{ \includegraphics[width=0.47\textwidth]{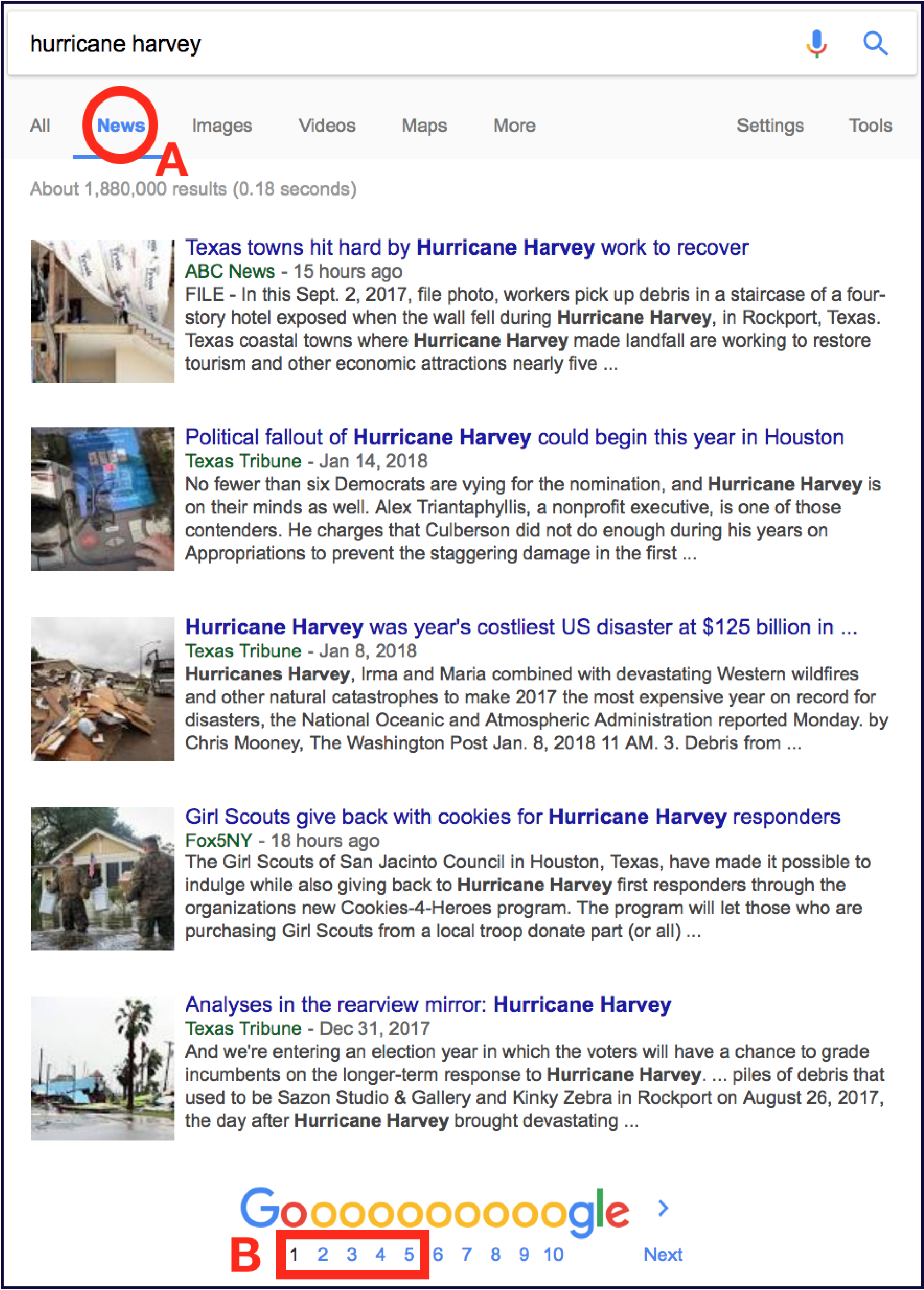} }}%
    \qquad
    \subfloat[A tweet from the Internet Archive requesting seeds for the U.S. Presidential Election collection.]{{ \includegraphics[width=0.39\textwidth]{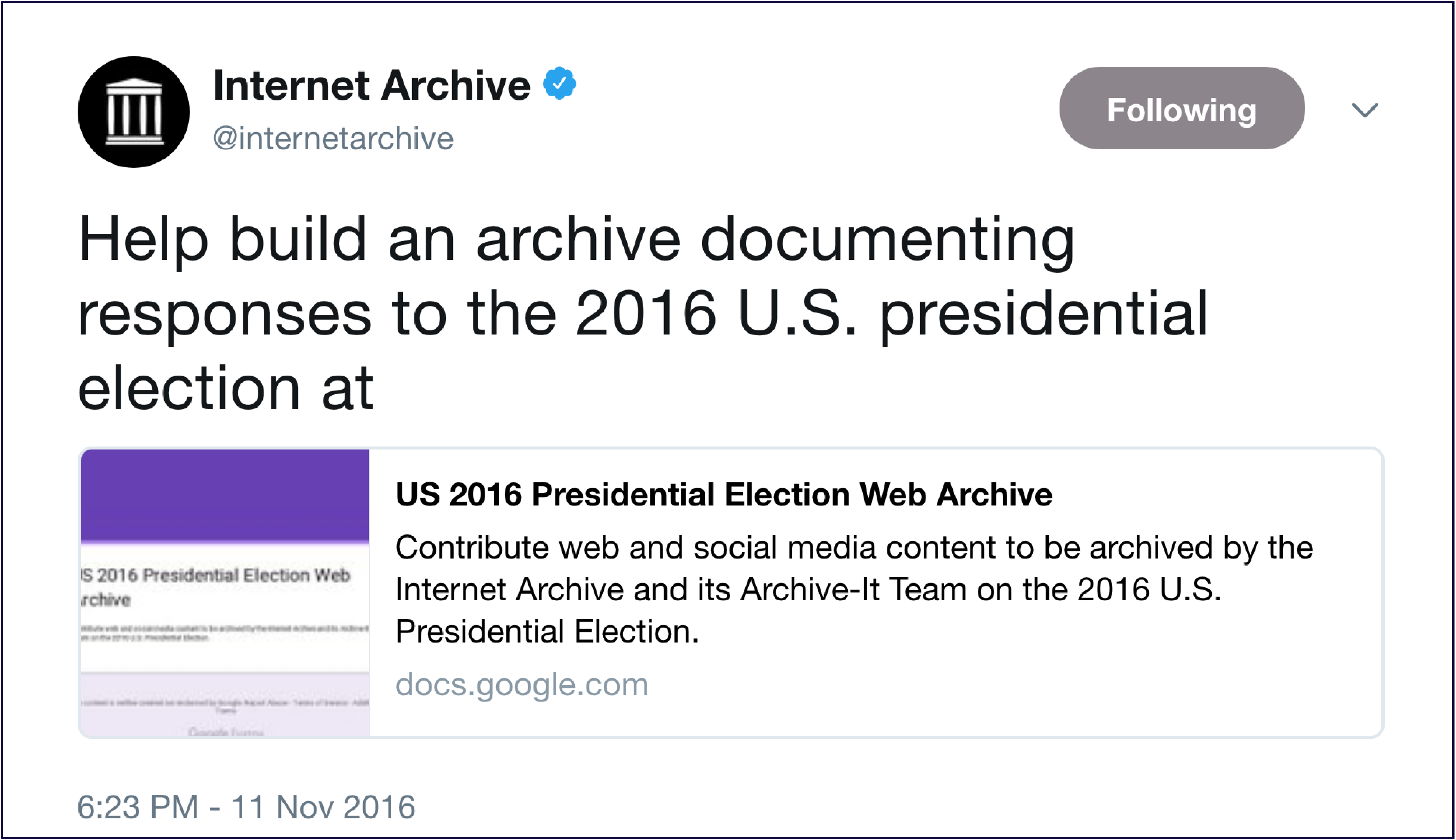} }}%
    \qquad
    \subfloat[A tweet from the Internet Archive requesting seeds for the Dakota Access Pipeline collection.]{{ \includegraphics[width=0.43\textwidth]{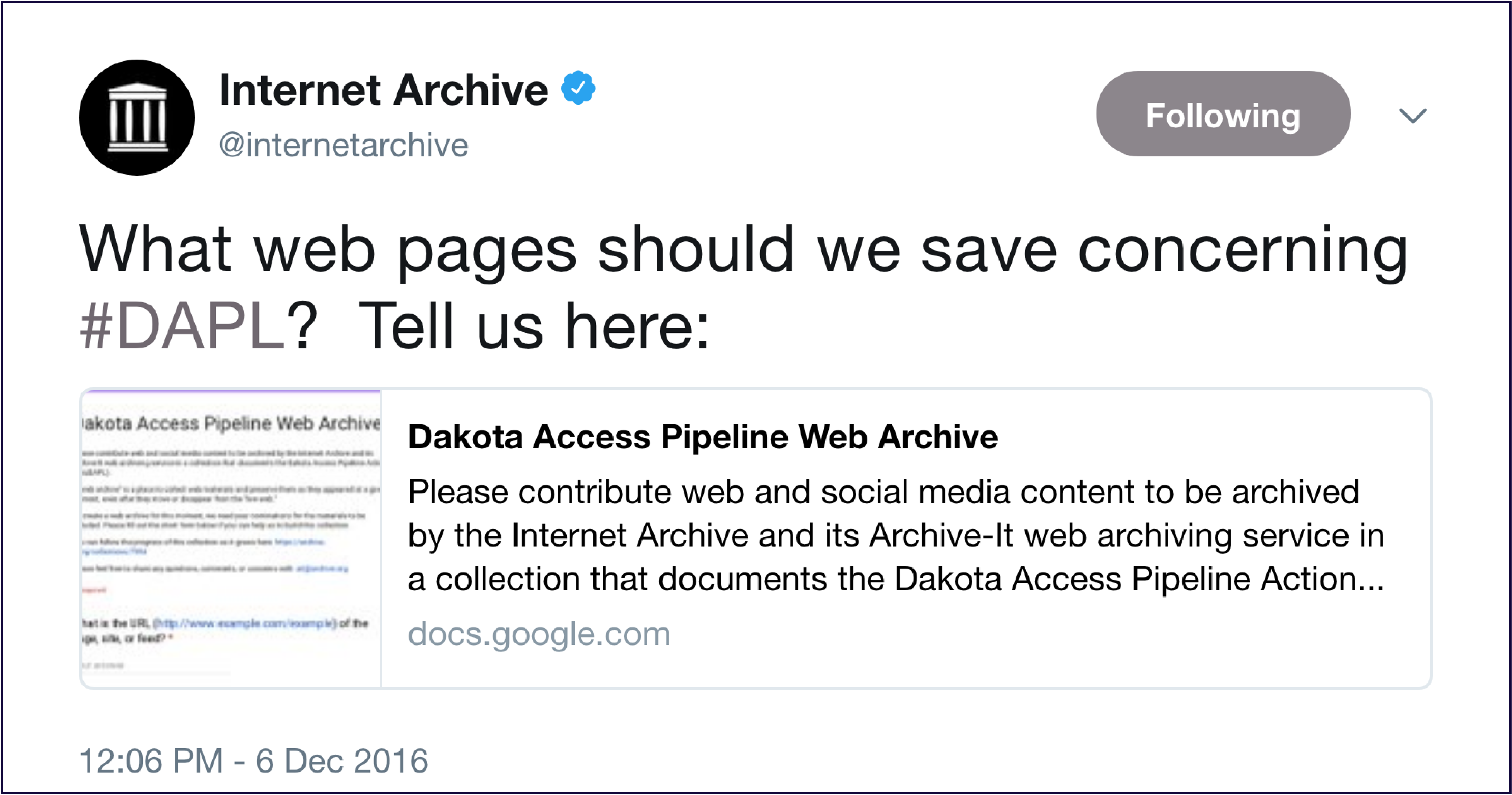} }}%
    \caption{a \& b: Google \textit{General} (a) and \textit{News} vertical (b) SERPs for query ``hurricane harvey,'' extracted 2018-01-16. Some links have been removed to enable showing more details. For the experiment, links were extracted from the first five pages (annotation b) of both SERPs for each query. c \& d: The Internet Archive has on multiple occasions requested that users submit seeds to bootstrap collections. The time when users respond with seeds impact the collections generated.}%
    \label{fig:SERPs}%
\end{figure*}

\begin{table*}
  \setlength{\tabcolsep}{1pt}
  \centering
  \caption{Rows 1 - 3: Sample collection of URIs extracted from Google on May 25, 2017 with query: ``healthcare bill.'' This shows the initial stages of the AHCA bill, highlighting the struggles to pass the bill. Rows 4 - 6: Sample collection of URIs extracted from Google on January 5, 2018 with query: ``healthcare bill.'' This shows the later (recent) stages of the AHCA bill, highlighting the failure of the bill which happened in September 2017.}
  \begin{tabular}{|c|c|c|} \hline 
  \textbf{\makecell{Publication Date}} & \textbf{Title} & \textbf{URI} \\ \Xhline{2\arrayrulewidth}
  
  2017-05-23 & \makecell{Healthcare saga shaping GOP approach\\ to tax bill} & \makecell{http://thehill.com/policy/finance/334650\\-healthcare-saga-shaping-gop-approach-to-tax-bill}\\ \hline
  2017-05-24 & \makecell{US Senate's McConnell sees tough\\path for passing healthcare bill} & \makecell{http://www.cnbc.com/2017/05/24/us-senates\\-mcconnell-sees-tough-path-for-passing-healthcare-bill.html} \\ \hline
  2017-05-25 & \makecell{Will the Republican Health Care\\Bill Really Lower Premiums?} & \makecell{http://time.com/4794400/health-care-premiums/} \\ \Xhline{2\arrayrulewidth}
  
  2017-12-19 & \makecell{House Republicans used lessons from failed\\health care bill to pass tax reform, Ryan says} & \makecell{https://www.pbs.org/newshour/politics/house-republicans-used\\-lessons-from-failed-health-care-bill-to-pass-tax-reform-ryan-says} \\ \hline
  2017-12-19 & \makecell{GOP tax bill also manages to\\needlessly screw up the healthcare system} & \makecell{http://www.latimes.com/business/lazarus/la-fi-\\lazarus-republican-tax-bill-individual-mandate-20171219-story.html} \\ \hline
  2017-12-20 & \makecell{How GOP tax bill's Obamacare changes\\will affect health care and consumers} & \makecell{http://www.chicagotribune.com/business/\\ct-biz-obamacare-insurance-penalty-repeal-1221-story.html}\\ \hline

  \end{tabular}
  \label{tab:ahcaInitial}
\end{table*}
From elections to natural disasters, web collections provide a critical source of information for researchers studying important historical events. Collections can be built automatically with focused crawlers or manually by an expert user. For example, an archivist at the National Library of Medicine collected seeds on Archive-It for the \textit{2014 Ebola Outbreak} event \cite{ChristieEbolaVirus}. Collections may also be built by multiple users. For example, the Internet Archive has on multiple occasions requested (Fig. \ref{fig:SERPs}c \& d) that users submit seeds via Google Docs to build collection for events such as the \textit{2016 US Presidential Election} and the \textit{Dakota Access Pipeline (DAPL)} event. Depending on when users begin to contribute seeds, URIs (Uniform Resource Identifiers) for early news stories may be difficult to discover via Google after one month, for as we show in this paper they can quickly fall to distant SERPs (Search Engine Result Pages).

Collection building often begins with a simple Google search to discover seeds. This can be done by issuing queries to Google and extracting URIs from the SERP (Fig. \ref{fig:SERPs}a \& b). For example, the following are two possible URI candidates extracted from the Google SERP (Fig. \ref{fig:SERPs}a) to include in a collection (or seed list) about the \textit{Hurricane Harvey} (August, 2017) event:
\small
\begin{verbatim}
http://www.cnn.com/specials/us/hurricane-harvey
https://www.nasa.gov/feature/goddard/2017/harvey-atlantic-ocean
\end{verbatim}
\normalsize
A SERP provides an opportunity to generate collections for news stories and events, therefore we focused on assessing the discoverability of news stories on the Google SERP. Queries used to extract news stories are examples of \textit{informational queries} \cite{broder2002taxonomy}, and we expect their SERP results to change as the news event evolves. We expect the SERP results for \textit{transactional} (e.g., ``samsung galaxy s3'') or \textit{navigational} (e.g., ``youtube'') to be less transient \cite{kim2011analysis}, but such queries are not the focus of our collection building effort. The URIs extracted from Google can serve as seeds: the seeds can be crawled to build collections in Archive-It, such as the Archive-It \textit{2017 Hurricane Harvey} collection\footnote{https://archive-it.org/collections/9323}. It is important to understand the behavior of Google as this will influence the collections or seeds generated from it. This is not easy because the inner workings of Google are proprietary, making it a black box. To build a representative collection about an event, it is important to capture not just a slice of time, but the various stages of the events \cite{risse2014you} - oldest to the newest. For example, on May 25, 2017, we issued the query: ``healthcare bill'' to Google and extracted links (Table \ref{tab:ahcaInitial}, 2017-05-23 -- 2017-05-25) from the SERP. Seven months later (January 5, 2017) we repeated the same operation (Table \ref{tab:ahcaInitial}, 2017-12-19 -- 2017-12-20). The May 2017 \textit{healthcare bill} collection shows the initial stages of the American Health Care Act of 2017 (AHCA) by highlighting the challenges facing the passage of the bill. On the other hand, the January 2018 collection shows a different (more recent) stage of the bill, highlighting the failure of the bill and the continued existence of Obamacare. These reflect the tendency of Google to highly rank newly created URIs for these kinds of queries. We quantify this behavior by estimating the rate at which new stories occur on the Google SERP.
The tendency of Google to return fresh documents can be altered by setting the custom date range parameter on the site. However, the date range information is not always available for the collections we intend to build. We explore how this parameter affects the kinds of documents returned. It is crucial to know the dynamics of finding initial stories on Google, as this would inform the time a collection building task ought to begin; if we know how fast new documents replace old documents on the Google SERP, we can plan collection building to ensure that URIs of older stories are included in the collection and not just the recent ones.

Accordingly, we conducted a longitudinal experiment to understand whether it is possible to retrieve a given URI of a news story over time, to gain insight about the appearance/disappearance of news stories across the pages in Google, and to identify the probability of finding the same URI of a story using the same query over time. This was achieved by issuing seven queries (Table \ref{tab:dataset}) every day for over seven months (2017-05-25 to 2018-01-12), and collecting links within \textit{h3} HTML tags from the first five pages (Fig. \ref{fig:SERPs}a \& b, annotation B) of the Google SERPs (Fig. \ref{fig:SERPs}a \& b). The queries were issued semi-automatically using a variant of the Local Stories Collection Generator \cite{chromeExtension}.

The longitudinal study resulted in the following contributions that shed some light on the discoverability and behavior of news stories on Google as time progresses. First, the tendency of Google to replace older documents with new ones is well known, but not the rate at which it occurs, our findings quantify this. Given two time intervals, e.g, day 0 and day 1, if we collected a set $x$ stories on day 0 and a set $y$ stories on day 1, the story replacement rate on day 1 is the fraction of stories found on day 0 but not found on day 1 ($\frac{|x - y|}{|x|}$). The daily story replacement rate on the Google \textit{General} SERP ranged from 0.21 -- 0.54, the weekly rate ranged from 0.39 -- 0.79, and monthly - 0.59 -- 0.92.  This means if you re-issued a query after one day, between 21\% to 54\% stories are replaced by new stories. But if you waited for one month, and issued the same query, between 59\% and 92\%  of the original stories are replaced. The \textit{News} vertical SERP showed a higher story replacement rates: daily - 0.31 -- 0.57, weekly - 0.54 -- 0.82, and monthly - 0.76 -- 0.92. Second, the probability of finding the same URI of a news story with the same query declines with time. For example, the probability of finding the same URI with the same query after one day (from the initial appearance on the \textit{General} SERP) is between 0.34 -- 0.44. This probability drops rapidly after a week, to a value between 0.01 -- 0.11. The probability is less for the \textit{News} vertical (daily - 0.28 -- 0.40, weekly - 0.03 -- 0.14, and approximately 0 one month later). We provide two predictive models that estimate the probabilities of finding an arbitrary URI of a news story on the \textit{General} and \textit{News} vertical SERPs as a function of time. Third, we show that news stories do not gradually progress from page 1 to page 2, 3, etc., and then out of view (beyond the page length considered). The progression we observed is less elegant (e.g., page 5 to 1, 3 to 2). These findings are highly informative to collection building efforts that originate from Google. For example, the results suggest that collections that originate from Google should begin days after an event happens, and should continue as time progresses to capture the various stages in the evolution of the event. Our research dataset comprising of 33,432 links extracted from the Google SERPs for over seven months, as well as the source code for the application utilized to semi-automatically generate the collections, are publicly available \cite{jcdl2018Repo}.

\section{Related work}
Since Chakrabarti et al. first introduced the focused crawler \cite{chakrabarti1999focused} as a system of discovering topic-specific web resources, there have been many research efforts pertaining to the generation of collections with some variant of a focused crawler. Bergmark \cite{bergmark2002collection} introduced a method for building collections by downloading web pages and subsequently classifying them into various topics in science, mathematics, engineering and technology. Farag et al. \cite{farag2017focused} introduced the Event Focused Crawler, a focused crawler for events which uses an event model to represent documents and a similarity measure to quantify the degree of relevance between a candidate URI and a collection. Klein et al. \cite{kleinWebSci} demonstrated that focused crawling on the archived web results in more relevant collections than focused crawling on the live web for events that occurred in the distant past. In order to augment digital library collections with publications not already in the digital library, Zhuang et al. \cite{zhuang2005s} proposed using publication metadata to help guide focused crawlers towards the homepages of authors. Klein et al. \cite{klein2007oai} also proposed a means of augmenting a digital library collection (the NASA Langley Research Center Atmospheric Science Data Center) with information discovered on the web using search engine APIs. SERPs are useful artifacts in their own right, and can be used for activities such as classifying queries as ``scholarly'' or ``non-scholarly'' \cite{nwala2016supervisedJCDL}. This work is similar to these efforts that use focused crawlers as it relates to collection building and using search engines to discover seeds, but we do not use a focused crawler to discover URIs.

Zheng et al. \cite{zheng2009graph} demonstrated that seed selection for crawlers is not a trivial problem because different seeds may result in collections that are considered ``good'' or ``bad,'' and proposed different seed selection algorithms. Similarly, Schneider et al. \cite{schneider2003building} expressed the difficulty in identifying seed URIs for building thematic collections, and suggested the continuous selection of seeds for collections about unfolding events such as the 9/11 attacks, to accommodate the evolution of the events. Baroni et al. \cite{baroni2009wacky} presented their findings about the effectiveness of various seed selection strategies as part of a broader effort to build a large linguistically processed web-crawled corpora. They demonstrated the discovery of a variety of seeds by issuing random queries to search engine APIs. Similar to these efforts, we consider seed selection a vital part of collection building but mainly focus on the temporal considerations when selecting seeds from search engines.

Cho and Garcia-Molina \cite{cho1999evolution} downloaded 720,000 pages (from 270 web servers) daily for four months to quantify the degree of change of web pages over time. They found that about 40\% of all web pages changed within a week, and 50\% of all the pages changed in about 50 days. Fetterly et al. \cite{fetterly2003large} extended Cho and Garcia-Molina's work by downloading (once a week for 11 weeks) over 150 million web pages and assessing the degree of change of web pages. They found that the average degree of change varies significantly across top-level domains, and that larger pages changed more frequently than smaller pages. Ntoulas et al. \cite{ntoulas2004s} focused on the evolution of link structure over time, the rate of creation of new pages, etc. They found high birth and death rates of pages with an even higher birth/death rates of the hyperlinks that connect the pages. 

In the web archiving community, link rot and content drift \cite{klein2014scholarly, klein2016scholarly} are two major reasons for collection building. Comparably, the difficultly in refinding news stories on the SERP suggests instant and persistent collection building by efforts that rely on the SERP for seed extraction or collection generation. McCown and Nelson \cite{mccown2007agreeing} issued queries for five months to search engine web user interfaces (Google, MSN and Yahoo) and their respective APIs, and found significant discrepancies in the results found on both interfaces. They also showed how search results decay over time and modeled the decay rate. Kim and Vitor \cite{kim2011analysis} studied Google, Bing, and Yahoo search engines, and showed that the top 10 results of 90\% of their queries were altered within ten days. There are many more studies that examine the evolution of web pages \cite{cho2003estimating, olston2008recrawl, adar2009web} or the web \cite{brewington2000dynamic, van2016estimating}. Our study is specific to news stories found in SERPs and not the evolution of the pages themselves. We sought to find out whether we could retrieve the same URI with the same query over a given period of time, instead of assessing the evolution of the content of individual web pages. This discoverability information is critical to collection building systems that utilize search engines. We tracked individual news stories to find when they were replaced by newer stories and quantified the rate at which older stories were replaced by newer stories. In addition to quantifying the rate of new stories as a function of time, we also quantify the rate of new stories for individual SERP pages. For example, we found that higher numbered pages (e.g., pages 3 - 5) have higher rates of new stories than lower numbered pages on Google (e.g., pages 1 - 2). Our results enable understanding the dynamics of refinding news stories on SERPs and are relevant to efforts that utilize search engines to discover seeds or to build collections.

Teevan et al. \cite{teevan2007information} illustrated the tendency of users to seek to refind web pages previously seen at a rate of about 40\% of all queries, and demonstrated how changes in search engine results can impede refinding links. Aula et al. \cite{aula2005information} also studied the prevalence of the ``re-find'' (re-access) behavior in a group 236 experienced web users, by investigating the problems with searching and re-accessing information. Their survey showed that the users often used some strategies such as using several browser windows concurrently to facilitate re-access. Capra et al. \cite{capra2005using} proposed a search engine use model as part of an effort to provide information to help better understand  how users find, refind, and manage information on the web. There are many other studies outlining attempts to refind web resources on search engines, such the effort of Bainbridge et al. \cite{bainbridge2017writers} to refind four copies of their published research papers on Google Scholar and ACM Digital Libary. Our research is similar to these efforts in the sense that we are interested in quantifying the likelihood of refinding URIs of news stories over time. However, the goal is not to fulfill the informational needs of particular users, but as part of an effort to extract seeds or generate collections from SERPs. Also, it is important to note that we utilize the search engine by issuing queries, it is not a known item search - we do not search for specific URIs to include in the collections we build - we let the SERP give us seeds. Instead we wish to know the rate as which the SERP produces the same URIs for the same query. Knowing the rate of new stories on the SERP for the same query indicates the ability of our collection generation process to refind previous stories and the rate at which the new stories from the SERP are included in the collection.
\section{Research questions}
Before employing search engines as seed generators we sought to first assess the discoverability of news stories on SERPs (top five pages), and understand the dynamics of refinding news stories on SERPs. Our ability to build representative collections for events from SERPs is tied to the ability of retrieving old and new stories from SERPs. Our primary research question was: what is the rate at which new stories replace old stories on the SERP over time? This rate information may provide the means to approximate our ability to refind older stories about an event and the rate at which the collection we generate from SERPs receives new stories from SERPs. Similarly, if we found a URI of a news story $s_0$ on day $d_0$, what is the probability that we would find $s_0$ on the next day $d_1$, or one week later on $d_7$? For example, if we found a URI for a news story on the SERP on Monday, what is the probability that we would refind the same URI with the same query on Tuesday (one day later) or next Monday (one week later)?

The generalization of the first main question led to the second - what is the rate of new stories for the individual SERPs or their pages? For example, does the \textit{General} SERP (or page 1) possess a higher new story rate than the \textit{News} vertical SERP (or page 2)? The pages with the lowest rate of new stories may indicate the page with the highest probability of finding the initial reports of an event. Understanding the characteristics of both SERP classes could inform the choice of SERPs when extracting seeds.

Finally, we sought to gain insight about how news stories on SERPs moved across the different pages.

\section{Methodology}
Here we describe the generation of our dataset, the measures we extracted from the dataset, and how these measures informed our research questions.

\begin{figure}
   \centering
   \includegraphics[width=0.47\textwidth]{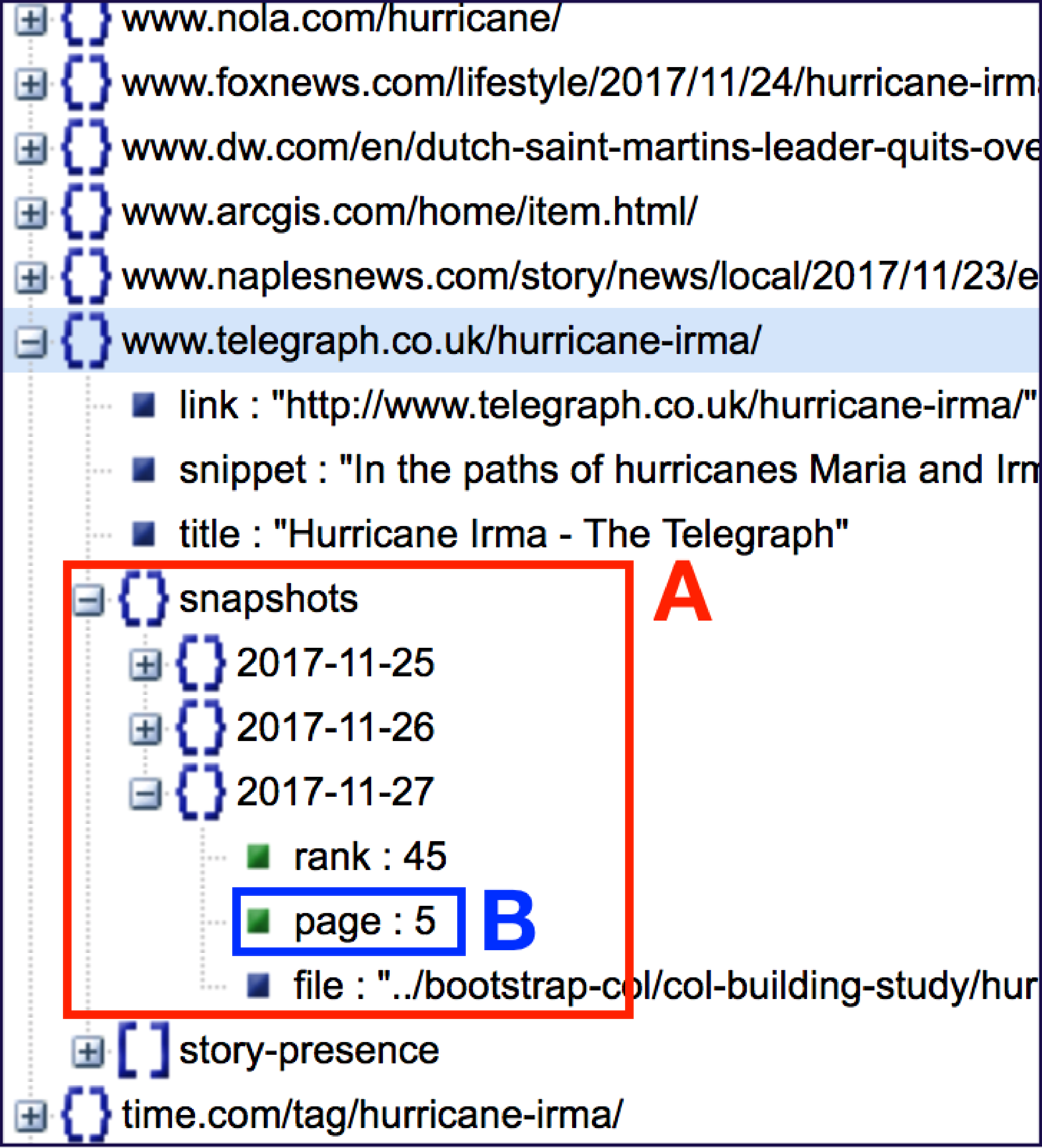}
   \caption{Representation of URIs collected from SERPs stores dates and the pages (1-5) the URIs were found.}
   \label{fig:repDS}%
\end{figure}
\subsection{Dataset generation, representation, and processing}
Seven queries representing public interest stories were selected: ``healthcare bill,'' ``manchester bombing,'' ``london terrorism,'' ``trump russia,'' ``travel ban,'' ``hurricane harvey,'' and ``hurricane irma.'' These queries represent various events that happened (or are happening) in different timelines. Consequently, the dataset extraction duration varied for the queries as outlined by Table \ref{tab:dataset}. The dataset extraction process lasted from 2017-05-25 to 2018-01-12. For each query, we extracted approximately 50 links within \textit{h3} HTML tags from the first five pages of the Google SERP from the default (\textit{All}) and \textit{News} vertical SERPs (Fig. \ref{fig:SERPs} a \& b). To avoid confusion, in this research we renamed the \textit{All} SERP to \textit{General} SERP. The first five pages were considered in order to gain better insight about the rate of new stories across pages, as considering a few pages (e.g., 1 or 2) may present an incomplete view. In total, 73,968 (13,708 unique) URIs were collected for the \textit{General} SERP and 77,634 (19,724 unique) for the \textit{News} vertical SERP (Table \ref{tab:dataset}).
\small
\begin{table}[h!]
    \setlength{\tabcolsep}{0.8pt}
   \centering
   \caption{Dataset generated by extracting URIs from SERPs (\textit{General} and \textit{News} vertical) for seven queries between 2017-05-25 and 2018-01-12.}
   \begin{tabular}{|l|c|c|c|c|c|}
          \hline
          \textbf{\makecell{Collection\\(Query/\\Topic)}} & \textbf{\makecell{Start date\\(duration\\in days)}} & \multicolumn{2}{c|}{\textbf{News story count}}  \\ \hline
          &  & \textbf{\makecell{General SERP count\\(unique count)}} & \textbf{\makecell{News vertical SERP count\\(unique count)}}  \\ \hline
          \makecell{healthcare\\bill}     & \makecell{2017-05-25\\(232)} & 12,809 (2,559) & 13,716 (3,450) \\ \hline
          \makecell{manchester\\bombing}  & \makecell{2017-05-25\\(232)} & 12,451 (1,018) & 13,751 (1,799) \\ \hline
          \makecell{london\\terrorism}    & \makecell{2017-06-04\\(222)} & 10,698 (1,098) & 10,450 (2,821) \\ \hline
          \makecell{trump\\russia}        & \makecell{2017-06-06\\(220)} & 12,311 (4,638) & 13,728 (3,482) \\ \hline
          \makecell{travel\\ban}          & \makecell{2017-06-07\\(219)} & 12,830 (2,849) & 13,439 (2,815) \\ \hline
          \makecell{hurricane\\harvey}    & \makecell{2017-08-30\\(135)} & 6,666 (685) & 6,450 (2,530) \\ \hline
          \makecell{hurricane\\irma}      & \makecell{2017-09-07\\(127)} & 6,203 (861) & 6,100 (2,827) \\ \hline \hline

          \multicolumn{2}{|r|}{\textbf{Subtotal}} & 73,968 (13,708) & 77,634 (19,724)  \\ \hline
          \multicolumn{2}{|r|}{\textbf{Collections Total}} & \multicolumn{2}{c|}{151,602 (33,432)} \\ \hline
          
   \end{tabular}
   \label{tab:dataset}
\end{table}
\normalsize

In previous work with the Local Memory Project (LMP) \cite{nwala2017local}, we introduced a local news collection generator \cite{chromeExtension}. The local news collection generator utilizes Google in order to build collections of stories from local newspapers and TV stations for US and non-US news sources. Unlike LMP, in this work we did not restrict the sources sampled to local news organization, but still utilized Google in order to discover seeds. The local news collection generator was used to scrape links from the Google SERP, and it was adapted to include the ability to extract all kinds of news stories from Google (not just from local news organizations). The Google search interface is meant for humans and not for robots and it presents a CAPTCHA when it is used too frequently in order to discourage automated searches. Consequently, the dataset collections were all generated semi-automatically with the use of the local news collection generator. The input provided to the extension was the query and the maximum number of pages to explore (five), and the output was a collection of URIs extracted from the SERPs.

The URIs collected daily from the SERPs were represented as JSON files. For a single query, two JSON files per day were generated, each file represented the URIs extracted from the \textit{General} SERP and \textit{News} vertical SERP. This means for a given day, a total of 14 (two per query) JSON files were generated. Each URI in a JSON file included  metadata extracted from the SERP such as the \textit{page number} and the \textit{rank} which is the position across all SERP pages (Fig. \ref{fig:repDS}). Additionally, each file included the date the data was generated.

At the center of the analysis was the ability to track the URI of a news story over time. A URI is a unique identifier for a resource, however, URIs often have aliases (multiple URIs identifying the same resource). For example, the following pair of URIs identify the same resource:

\small
\begin{verbatim}
(a) https://www.redcross.org/donate/disaster-donations?campname=
irma&campmedium=aspot
(b) https://www.redcross.org/donate/disaster-donations
\end{verbatim}
\normalsize
As a result, we transformed all URIs before matching by trimming the scheme and all parameters from the URIs, using a method suggested by Brunelle et al. \cite{brunelle2015archiving}. The parameters in URIs often express a reference source such as \texttt{origin} and \texttt{callback}, or session parameters such as \texttt{session}. The transformed version of the URI was used to track the individual news stories. Subsequently, for each news story we recorded all the dates and pages it was observed on the SERP. For example, Fig. \ref{fig:repDS} shows that the URI ``http://www.telegraph.co.uk/hurricane-irma/'' was observed between 2017-11-25 to 2017-11-27 (Fig. \ref{fig:repDS} annotation A), and was extracted from page 5 on 2017-11-27 (Fig. \ref{fig:repDS} annotation B).

\subsection{Primitive measures extraction}
\label{sec:primitiveMeasures}
The following measures were extracted from the dataset and provided information to help answer our research questions.

\subsubsection{\textbf{Story replacement rate, new story rate, and page level new story rate}} \maskColor{.}\\

Given that at time point $t_0$ we observed a set of URIs for news stories $u_0$ and at time point $t_1$ we observed a set of URIs for news stories $u_1$, then the story replacement rate at $t_1$ is $\frac{|u_0 - u_1|}{|u_0|}$. For example, if we observed URIs  $\{a, b, c\}$ at $t_0$ and URIs $\{a, b, x, y\}$ at $t_1$, then the story replacement rate at $t_1$ is 

$\frac{|\{a, b, c\} - \{a, b, x, y\}|}{|\{a, b, c\}|} = \frac{|\{c\}|}{|\{a, b, c\}|} = \frac{1}{3} = 0.3.$ 
\\This means that at $t_1$ \textit{c}  was replaced. Similarly, the rate of new stories going from $t_0$ to $t_1$ is $\frac{|u_1 - u_0|}{|u_1|}$. For example, if we observed URIs $\{a, b, c\}$ at $t_0$ and URIs $\{a, b, c, d, e\}$ at $t_1$, then the new story rate from $t_0$ to $t_1$ is 

$\frac{|\{a, b, c, d, e\} - \{a, b, c\}|}{|\{a, b, c, d, e\}|} = \frac{|\{d, e\}|}{|\{a, b, c, d, e\}|} = \frac{2}{5} = 0.4.$
\\This means that at $t_1$ we observed new stories $d$ and $e$. We calculated the story replacement and new story rates using different temporal intervals (daily, weekly, and monthly) for the individual first five pages of the \textit{General} and \textit{News} vertical SERPs. The daily story replacement rate indicates the proportion of stories replaced on a daily basis. This is similar to the daily new story rate because the SERP returns a similar number of results ($mean = median = mode = $ 10 links, and $\sigma = 0.43$). The daily new story rate approximately indicates the rate of new stories that replaced previously seen stories on the SERP on a daily basis. The higher the story replacement and new story rates, the lower the likelihood of refinding previously seen stories.

\subsubsection{\textbf{Probability of finding a story}} \maskColor{.}\\
\label{sec:probFinding}

Given a collection of URIs for news stories for a topic (e.g., ``hurricane harvey''), consider the URI for a story $s_0$ that was observed for the first time on page 4 of the SERP on day $d_0$. We represent this as $s_0^{d_0} = 4$. If we find $s_0$ on page 2 on the next day $d_1$ and then it disappears for the next two days, we represent the timeline observation of $s_0$ as $\{4, 2, 0, 0\}$. Therefore, given a collection (e.g., ``hurricane harvey'') of $N$ URIs for news stories, the probability $P(s^{d_k})$ that the URI of a story $s$ is seen after $k$ days ($d_k$), is calculated using Eqn. \ref{eqn:probCalc0}.

\begin{equation}
    P(s^{d_k}) = \frac{\sum_{n=1}^{N} T(s_i^{d_k})}{N}
  ;
  T(s_i^{d_i}) = 
  \begin{cases}
       0 & \text{; if $s_i^{d_i} = 0$} \\
       1 & \text{; if $s_i^{d_i}> 0$} \\
  \end{cases}
  \label{eqn:probCalc0}
\end{equation}

The probability $P(s^{d_k} = m)$ that the URI of a story $s$ is seen after $k$ days ($d_k$) on page $m$, is calculated using Eqn. \ref{eqn:probCalc1}.

\begin{equation}
    P(s^{d_k} = m) = \frac{\sum_{n=1}^{N} T(s_i^{d_k})}{N}
  ;
  T(s_i^{d_i}) = 
  \begin{cases}
       0 & \text{; if $s_i^{d_i} \neq m$} \\
       1 & \text{; if $s_i^{d_i} = m$} \\
  \end{cases}
  \label{eqn:probCalc1}
\end{equation}

\subsubsection{\textbf{Distribution of stories over time across pages}} \maskColor{.}\\

For each story URI, we recorded the dates it was observed on the SERP. For each date, we recorded the page where the story was found. The collection of stories and the date/page observations were expressed using the notation introduced in Section \ref{sec:probFinding}. For example, the following list of three URIs for news stories $s_0, s_1,$ and $s_2$ were observed for the first time (first day - $d_0$) on pages, 4, 1, and 1, respectively. On the last day ($d_3$), the first story ($s_0$) was not seen on any of the pages ($s_0^{d_3} = 0$), however both the second ($s_1$) and third ($s_2$) stories were found on the first page ($s_1^{d_3} = 1$ and $s_2^{d_3} = 1$):

$s_0 = \{4, 2, 0, 0\}$, 

$s_1 = \{1, 2, 0, 1\}$, and

$s_2 = \{1, 1, 1, 1\}$.
\subsubsection{\textbf{Overlap rate and recall}} \maskColor{.}\\

Given two sets of collections of URIs, \textit{A} and \textit{B}, the overlap rate $O(A, B)$ quantifies the amount of URIs common within both sets without considering the size disparities of the sets. This was calculated using the Overlap coefficient as follows: $O(A, B) = \frac{|A \cap B|}{min(|A|, |B|)}$. The standard information retrieval recall metric $r(A, B)$ for two sets of collections \textit{A} and \textit{B} with respect to \textit{A}, quantifies the amount of stories present in \textit{A} and \textit{B} (as a fraction of \textit{A}) was calculated as $r(A, B) = \frac{|A \cap B|}{|A|}$.

Our dataset was generated without setting any parameters on the Google SERP. However, the Google SERP provides a date range parameter that attempts to restrict the documents returned on the SERP to documents published within the date range. For example, setting the date range to \textit{2017-06-01} and \textit{2017-06-30}, attempts to restrict the documents in the SERP to those published between June 1, 2017 and June 30, 2017. To assess the effect of setting the date range parameter on discovering older stories that fall within a specific timeframe, we took the following steps. First, from our original dataset, we selected five collections of stories for queries about topics that occurred before June 2017: ``healthcare bill,'' ``trump russia,'' ``travel ban,'' ``manchester bombing,'' and ``london terrorism.'' This set of five collections was called \textit{June-2017}. Second, we removed all stories from \textit{June-2017} that were not published in June 2017. Third, we issued the selected five queries to the Google SERP without setting the date range to generate five additional collections (from the first five pages). This set of five collection was called \textit{Jan-2018} (control test collection). Fourth, we issued the same five queries to the Google SERP, but this time, we set the date range to \textit{2017-06-01} and \textit{2017-06-30}, and extracted five collections. This set of five collections was called \textit{Jan-2018-Restricted-to-June}. Finally, we calculated the overlap rate and recall between the \textit{June-2017} and \textit{Jan-2018}, as well as \textit{June-2017} and \textit{Jan-2018-Restricted-to-June} collections for the pairs of collections with the same query.

\section{Results and Discussion}

\begin{figure*}
   \centering
   \includegraphics[width=\textwidth]{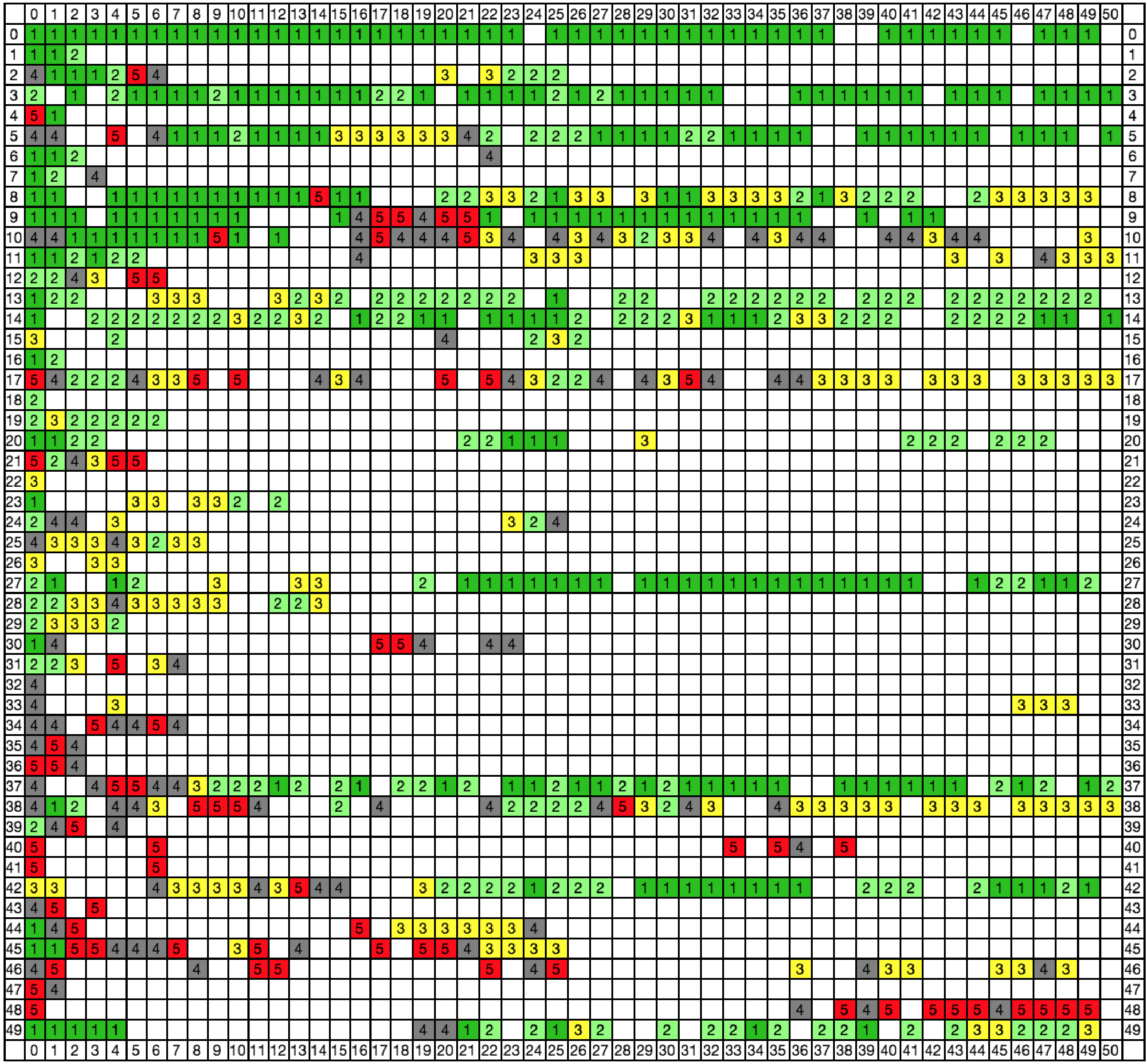}
   \caption{Page-level temporal distribution of stories in the ``manchester bombing'' \textit{General} SERP collection showing multiple page movement patterns. Stories in \textit{General} SERP collections persist longer than stories in \textit{News} vertical collections. Color codes - \textcolor[RGB]{34, 185, 4}{page 1}, \textcolor[RGB]{128, 255, 104}{page 2}, \textcolor[RGB]{230, 230, 0}{page 3}, \textcolor[RGB]{109, 109, 109}{page 4}, \textcolor[RGB]{251, 0, 6}{page 5}, and blank for outside pages 1 - 5.}
   \label{tab:normTempDistG}%
\end{figure*}

\begin{figure*}
   \centering
   \includegraphics[width=\textwidth]{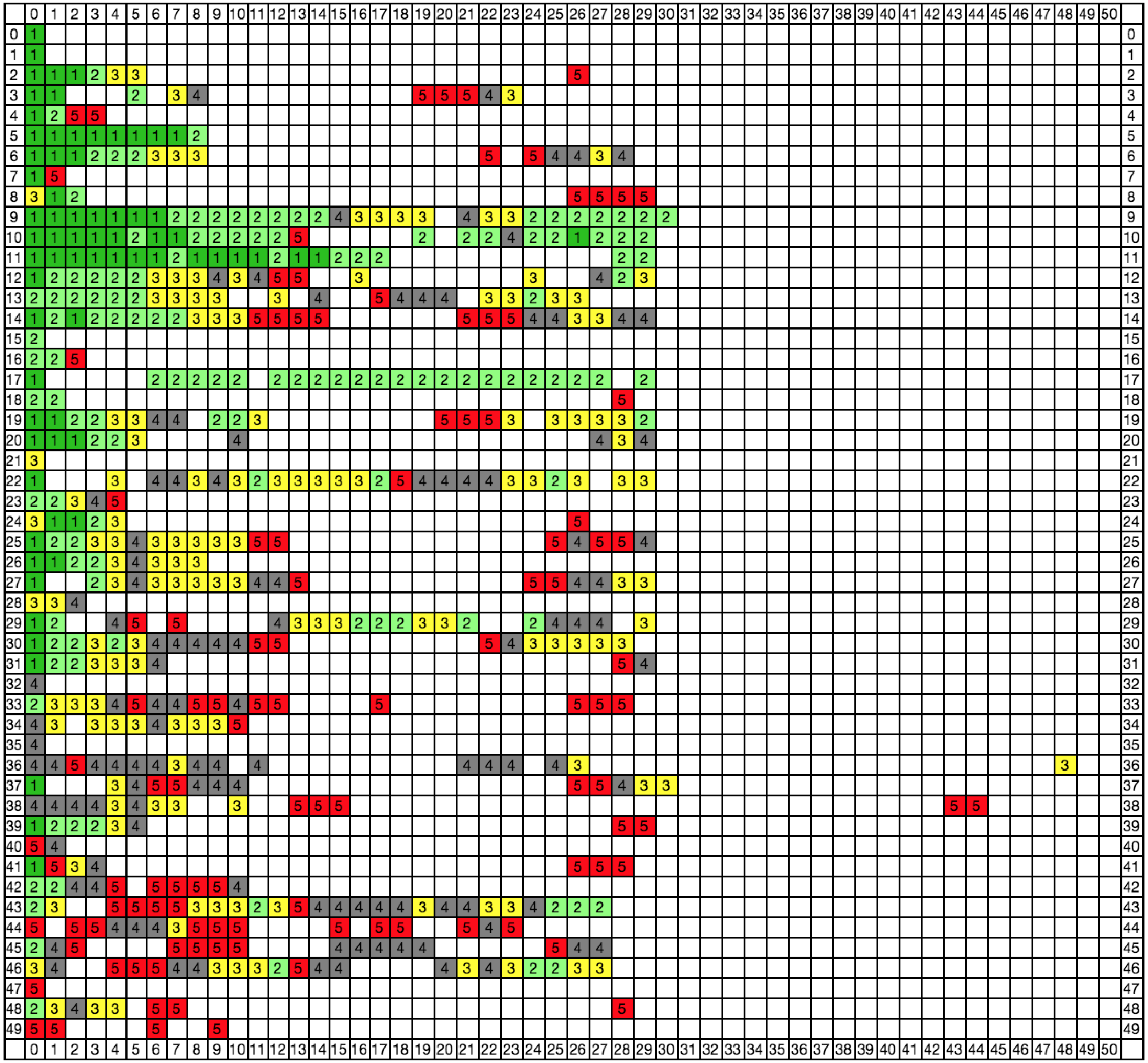}
   \caption{Page-level temporal distribution of stories in the ``manchester bombing'' \textit{News} vertical SERP collection showing multiple page movement patterns, and the shorter lifespan of \textit{News} vertical URIs (compared to \textit{General} SERP URIs). Color codes - \textcolor[RGB]{34, 185, 4}{page 1}, \textcolor[RGB]{128, 255, 104}{page 2}, \textcolor[RGB]{230, 230, 0}{page 3}, \textcolor[RGB]{109, 109, 109}{page 4}, \textcolor[RGB]{251, 0, 6}{page 5}, and blank for outside pages 1 - 5.}
   \label{tab:normTempDistNV}%
\end{figure*}

\begin{figure*}
  \subfloat[``hurricane harvey'' \textit{General} SERP collection]{{ \includegraphics[width=0.47\textwidth]{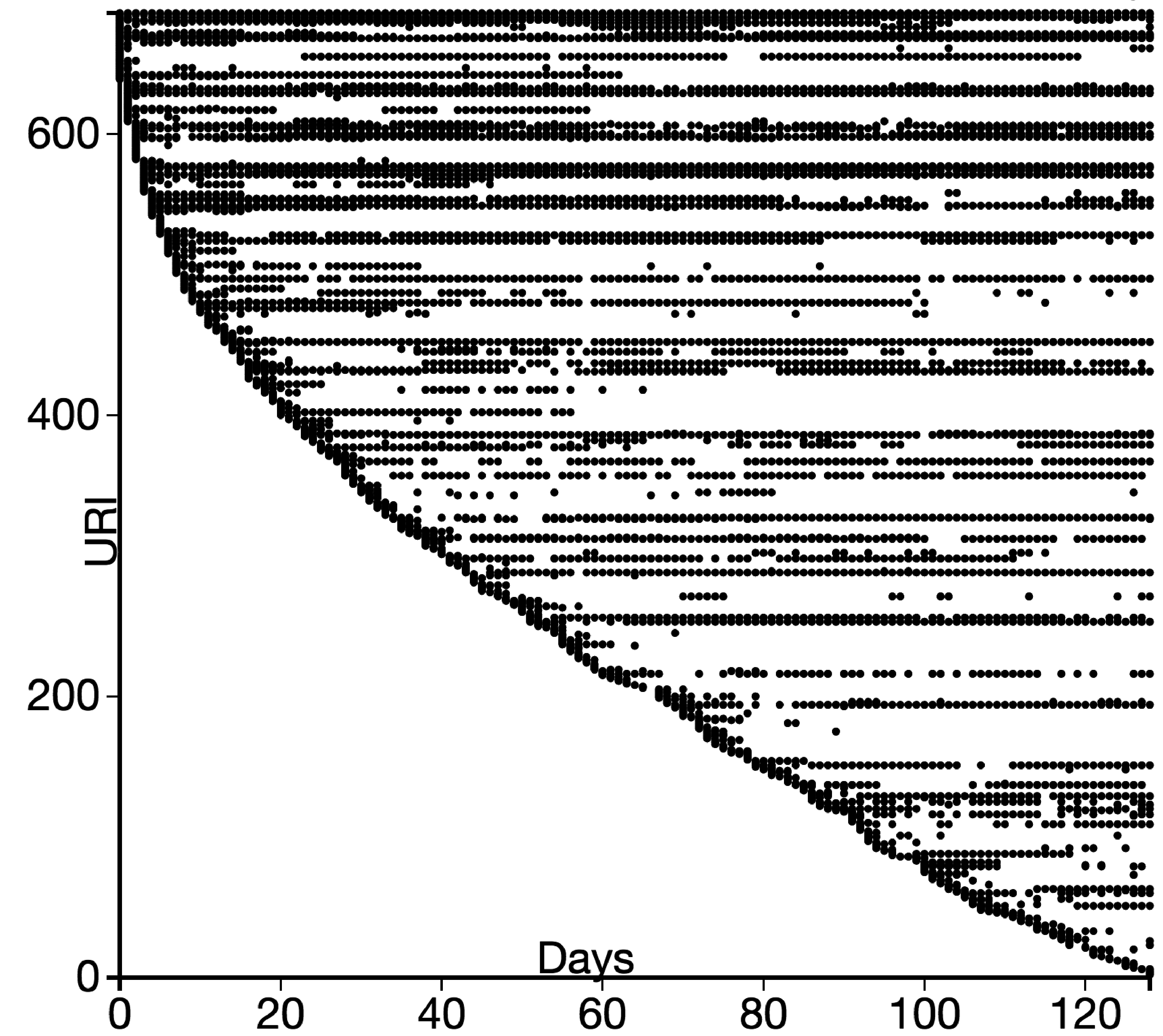} }}
  \qquad
  \subfloat[``hurricane harvey'' \textit{News} vertical SERP collection]{{ \includegraphics[width=0.47\textwidth]{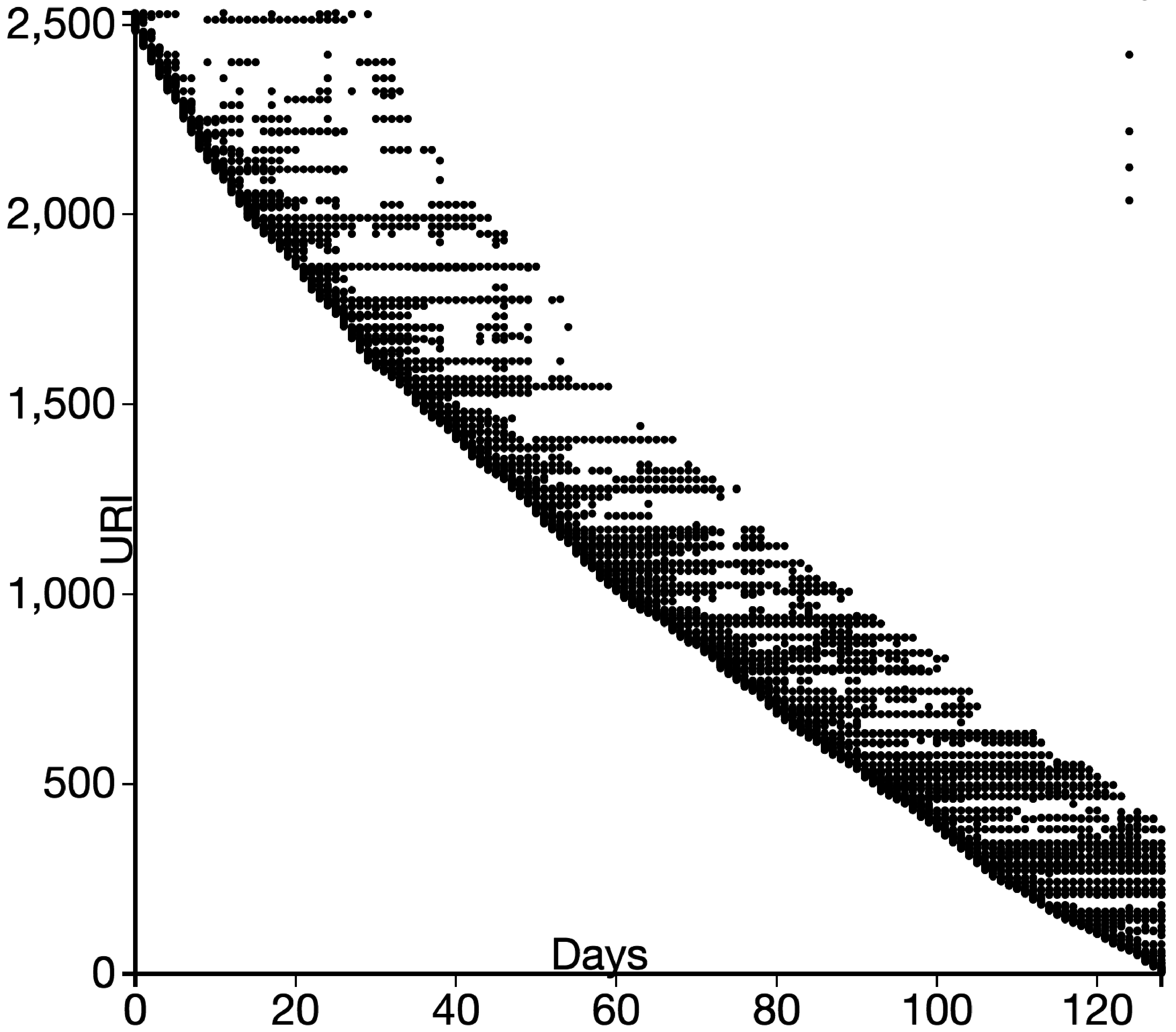} }}
  \qquad
  \subfloat[``trump russia'' \textit{General} SERP collection]{{ \includegraphics[width=0.47\textwidth]{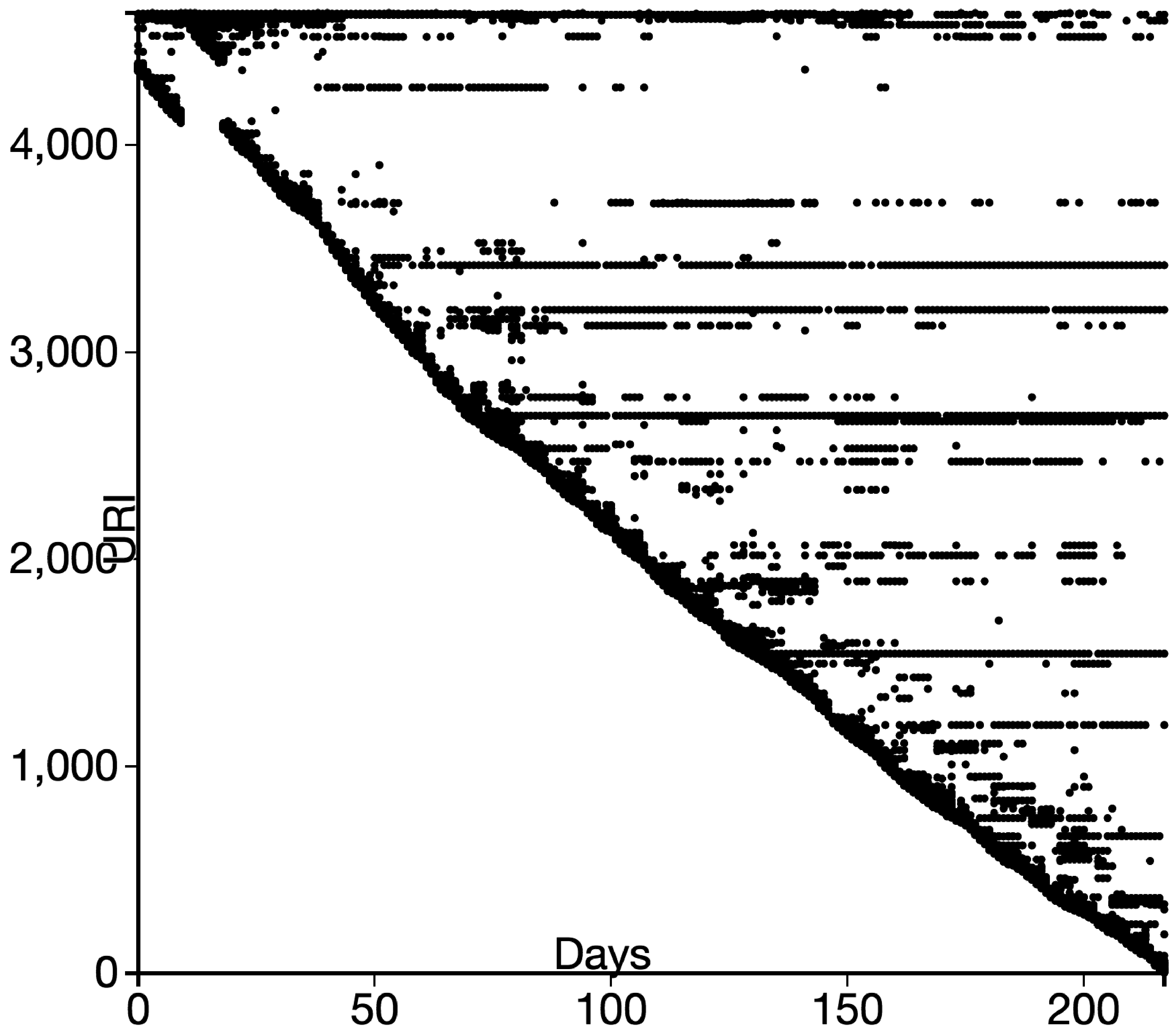} }}
  \qquad
  \subfloat[``trump russia'' \textit{News} vertical SERP collection]{{ \includegraphics[width=0.47\textwidth]{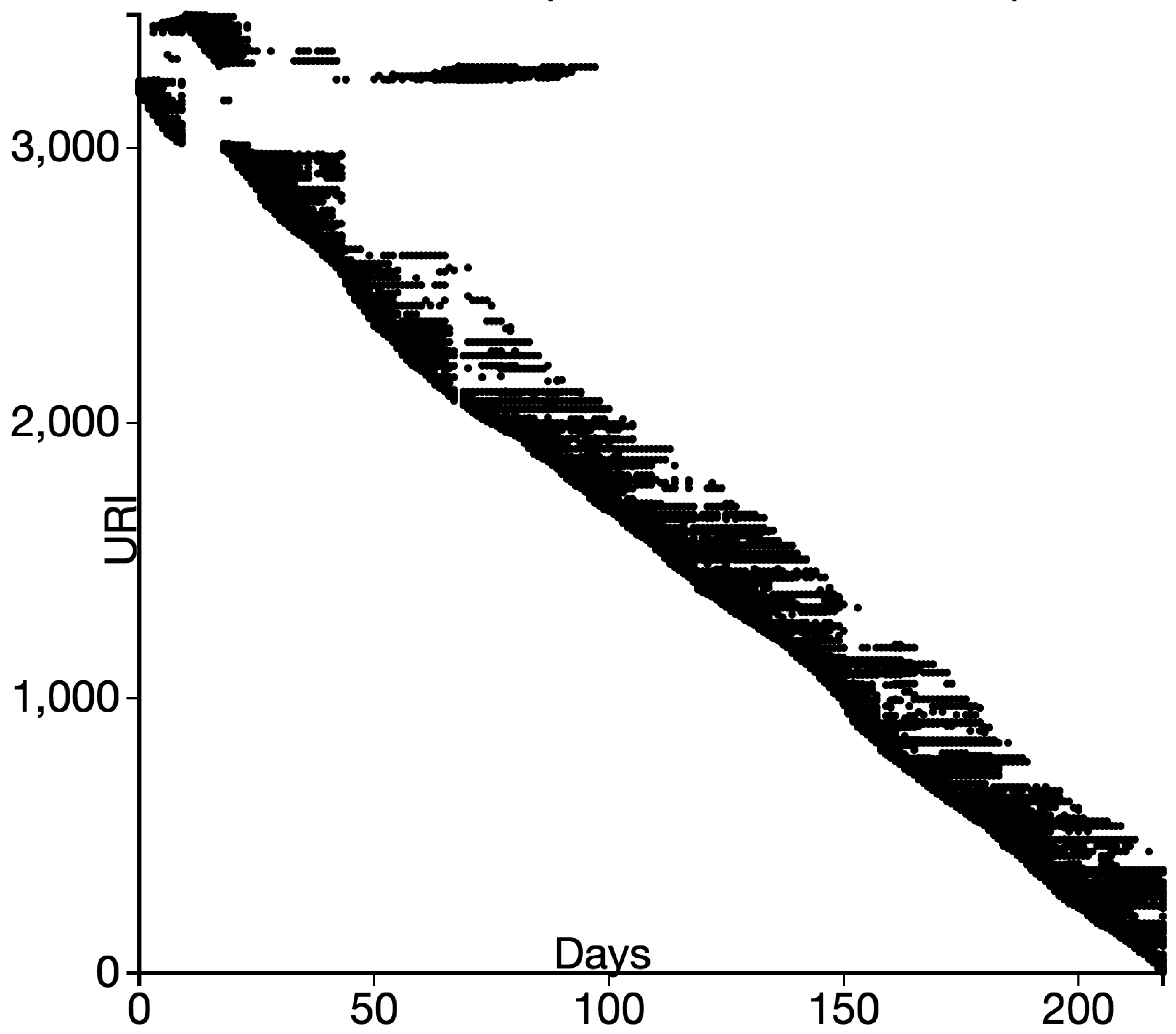} }}
  \caption{Temporal distributions: Stories in \text{General} SERP collections (a \& c) persist longer (``longer life'') than stories in \textit{News} vertical collections (b \& d). Compared to the ``trump russia'' \textit{General} SERP collection, the stories in the ``hurricane harvey'' \textit{News} vertical collection have a ``longer life'' due to a lower rate of new stories.}
  \label{tab:tempDist}
\end{figure*}
\small
\begin{table}
    \setlength{\tabcolsep}{0.8pt}
   \centering
   \caption{Average story replacement rate for \textit{General} and \textit{News} vertical SERP collections. Column markers: \minColor{ minimum$^{-}$ } and \maxColor{ maximum$^{+}$}.}
   \begin{tabular}{|l||c|c|c||c|c|c|}
          \hline
          \textbf{Collection} & \multicolumn{3}{c||}{\textbf{General SERP}} & \multicolumn{3}{c|}{\textbf{News vertical SERP}} \\ \hline
           & \textbf{Daily} & \textbf{Weekly} & \textbf{Monthly} & \textbf{Daily} & \textbf{Weekly} & \textbf{Monthly} \\ \hline
          healthcare bill     & 0.42 & 0.60 & 0.76 & 0.44 & 0.71 & 0.87  \\ \hline
          manchester bombing  & 0.27 & \minColor{ $\boldsymbol{0.39^{-}}$ } & \minColor{ $\boldsymbol{0.59^{-}}$ } & \minColor{ $\boldsymbol{0.31^{-}}$ } & \minColor{ $\boldsymbol{0.54^{-}}$ } & \minColor{ $\boldsymbol{0.76^{-}}$ } \\ \hline
          london terrorism    & 0.34 & 0.41 & 0.60 & 0.43 & 0.66 & 0.84 \\ \hline
          trump russia        & \maxColor{ $\boldsymbol{0.54^{+}}$ } & \maxColor{ $\boldsymbol{0.79^{+}}$ } & \maxColor{ $\boldsymbol{0.92^{+}}$ } & 0.42 & 0.71 & 0.90 \\ \hline
          travel ban          & 0.43 & 0.63 & 0.82 & 0.45 & 0.62 & 0.83 \\ \hline
          hurricane harvey    & \minColor{ $\boldsymbol{0.21^{-}}$ } & 0.41 & 0.67 & 0.49 & 0.77 & 0.91 \\ \hline
          hurricane irma      & 0.27 & 0.44 & 0.73 & \maxColor{ $\boldsymbol{0.57^{+}}$ } & \maxColor{ $\boldsymbol{0.82^{+}}$ } & \maxColor{ $\boldsymbol{0.92^{+}}$ } \\ \hline
          
   \end{tabular}
   \label{tab:avgStoryReplacementRate}
\end{table}
\begin{table}
    \setlength{\tabcolsep}{0.8pt}
   \centering
   \caption{Average new story rate for \textit{General} and \textit{News} vertical SERP collections. Column markers: \minColor{ minimum$^{-}$ } and \maxColor{ maximum$^{+}$}.}
   \begin{tabular}{|l||c|c|c||c|c|c|}
          \hline
          \textbf{Collection} & \multicolumn{3}{c||}{\textbf{General SERP}} & \multicolumn{3}{c|}{\textbf{News vertical SERP}} \\ \hline
           & \textbf{Daily} & \textbf{Weekly} & \textbf{Monthly} & \textbf{Daily} & \textbf{Weekly} & \textbf{Monthly} \\ \hline
          healthcare bill     & 0.42 & 0.58 & 0.62  &  0.44 & 0.70 & 0.82  \\ \hline
          manchester bombing  & 0.27 & \minColor{ $\boldsymbol{0.37^{-}}$ } & \minColor{ $\boldsymbol{0.46^{-}}$ }  & \minColor{ $\boldsymbol{0.31^{-}}$ } & \minColor{ $\boldsymbol{0.52^{-}}$ }& \minColor{ $\boldsymbol{0.66^{-}}$ } \\ \hline
          london terrorism    & 0.34 & 0.40 & 0.51  & 0.43 & 0.65 & 0.84 \\ \hline
          trump russia        & \maxColor{ $\boldsymbol{0.54^{+}}$ } & \maxColor{ $\boldsymbol{0.78^{+}}$ } & \maxColor{ $\boldsymbol{0.83^{+}}$ } &  0.42 & 0.70 & 0.83 \\ \hline
          travel ban          & 0.43 & 0.62 & 0.71 & 0.45 & 0.61 & 0.75 \\ \hline
          hurricane harvey    & \minColor{ $\boldsymbol{0.21^{-}}$ } & 0.38 & 0.51 & 0.49 & 0.76 & 0.82 \\ \hline
          hurricane irma      & 0.27 & 0.41 & 0.61 & \maxColor{ $\boldsymbol{0.57^{+}}$ } & \maxColor{ $\boldsymbol{0.81^{+}}$ } & \maxColor{ $\boldsymbol{0.91^{+}}$ } \\ \hline
          
   \end{tabular}
   \label{tab:avgNewStoryRate}
\end{table}

\begin{table}
    \setlength{\tabcolsep}{0.8pt}
   \centering
   \caption{Probability of finding the same story after one day, one week, and one month (from first observation) for \textit{General} and \textit{News} vertical SERP collections. Column markers: \minColor{ minimum$^{-}$ } and \maxColor{ maximum$^{+}$}.}
   \begin{tabular}{|l||c|c|c||c|c|c|}
          \hline
          \textbf{Collection} & \multicolumn{3}{c||}{\textbf{General SERP}} & \multicolumn{3}{c|}{\textbf{News vertical SERP}} \\ \hline
           & \textbf{a day} & \textbf{a week} & \textbf{a month} & \textbf{a day} & \textbf{a week} & \textbf{a month} \\ \hline
          healthcare bill     & 0.35 & 0.04 & 0.02 & 0.34 & 0.07 & 0.00   \\ \hline
          manchester bombing  & \maxColor{ $\boldsymbol{0.44^{+}}$ }  & 0.09 & 0.07 & \maxColor{ $\boldsymbol{0.40^{+}}$ }  & \maxColor{ $\boldsymbol{0.14^{+}}$ } & 0.00  \\ \hline
          london terrorism    & 0.37 & \maxColor{ $\boldsymbol{0.11^{+}}$ } & 0.07 & 0.34  & 0.09 & 0.00  \\ \hline
          trump russia        & 0.39 & \minColor{ $\boldsymbol{0.01^{-}}$ } & \minColor{ $\boldsymbol{0.01^{-}}$ } & 0.36  & 0.10 & 0.00  \\ \hline
          travel ban          & 0.43 & 0.06 & 0.02 & 0.32  & 0.12 & 0.00  \\ \hline
          hurricane harvey    & 0.38 & 0.10 & \maxColor{ $\boldsymbol{0.08^{+}}$ } & 0.29  & 0.05 & 0.00  \\ \hline
          hurricane irma      & \minColor{ $\boldsymbol{0.34^{-}}$ }  & 0.07 & 0.05 & \minColor{ $\boldsymbol{0.28^{-}}$ }  & \minColor{ $\boldsymbol{0.03^{-}}$ } & 0.00  \\ \hline
          
   \end{tabular}
   \label{tab:probFindingStory}
\end{table}
\normalsize
\small
\begin{table}
    \setlength{\tabcolsep}{1pt}
   \centering
   \caption{Comparison of two collections against the \textit{June-2017} collection (documents published in June 2017). The collection \textit{Jan-2018}, which was created (2018-01-11) without modifying the SERP date range parameter has a lower overlap than the collection (\textit{June-2018-Restricted-to-June}) created the same day (2018-01-11) by setting the SERP date range parameter to June 2017. Even though setting the date range parameter increases finding stories with common publication dates as the date range, the recall is poor due to the fixed SERP result. Column markers: \maxColor{ \textbf{maximum}}.}
   \begin{tabular}{|c||c||c|c|c||c|c|c|}
      \hline
      \textbf{Collection} & \textbf{Metrics} & \multicolumn{3}{c||}{\textbf{General SERP}} & \multicolumn{3}{c|}{\textbf{News vertical SERP}} \\ \hline
      &  & \textbf{\textit{\makecell{June\\-2017}}} & \textbf{\textit{\makecell{Jan\\-2018}}} & \textbf{\textit{\makecell{Jan-2018-\\Restricted-\\to-June}}} & \textbf{\textit{\makecell{June\\-2017}}} & \textbf{\textit{\makecell{Jan\\-2018}}} & \textbf{\textit{\makecell{Jan-2018-\\Restricted-\\to-June}}} \\ \Xhline{2\arrayrulewidth}
      \multirow{3}{*}{ \makecell{healthcare\\bill} } & size     &  460 & 51 & 50 & 419 & 50 & 50  \\ \cline{2-8}
      & overlap                                                 &  1.00 & 0.06  & \maxColor{ $\boldsymbol{0.60}$ } & 1.00 & 0.02 & \maxColor{ $\boldsymbol{0.56}$ } \\  \cline{2-8}
      & recall                                                  &  1.00 & 0.01  & \maxColor{ $\boldsymbol{0.07}$ } & 1.00 & 0.00 & \maxColor{ $\boldsymbol{0.07}$ } \\ \Xhline{2\arrayrulewidth}
      
      \multirow{3}{*}{ \makecell{manchester\\bombing} } & size  &  483 & 50 & 51 & 50 & 50 & 548  \\ \cline{2-8}
      & overlap                                                 &  1.00 & 0.04 & \maxColor{ $\boldsymbol{0.82}$ } & 1.00 & 0.00 & \maxColor{ $\boldsymbol{0.50}$ } \\  \cline{2-8}
      & recall                                                  &  1.00 & 0.00 & \maxColor{ $\boldsymbol{0.08}$ } & 1.00 & 0.00 & \maxColor{ $\boldsymbol{0.05}$ } \\ \Xhline{2\arrayrulewidth}
      
      \multirow{3}{*}{ \makecell{london\\terrorism} } & size    &  191 & 50 & 52 & 50 & 50 & 172  \\ \cline{2-8}
      & overlap                                                 &  1.00 & 0.09 & \maxColor{ $\boldsymbol{0.70}$ } & 1.00 & 0.00 & \maxColor{ $\boldsymbol{0.68}$ } \\  \cline{2-8}
      & recall                                                  &  1.00 & 0.02 & \maxColor{ $\boldsymbol{0.18}$ } & 1.00 & 0.00 & \maxColor{ $\boldsymbol{0.20}$ } \\ \Xhline{2\arrayrulewidth}
      
      \multirow{3}{*}{ \makecell{trump\\russia} } & size        &  562 & 50 & 51 & 50 & 50 & 524  \\ \cline{2-8}
      & overlap                                                 &  1.00 & 0.00 & \maxColor{ $\boldsymbol{0.54}$ } & 1.00 & 0.00 & \maxColor{ $\boldsymbol{0.58}$ } \\  \cline{2-8}
      & recall                                                  &  1.00 & 0.00 & \maxColor{ $\boldsymbol{0.05}$ } & 1.00 & 0.00 & \maxColor{ $\boldsymbol{0.06}$ } \\ \Xhline{2\arrayrulewidth}
      
      \multirow{3}{*}{ \makecell{travel\\ban} } & size          &  391 & 50 & 52 & 50 & 50 & 370  \\ \cline{2-8}
      & overlap                                                 &  1.00 & 0.04 & \maxColor{ $\boldsymbol{0.84}$ } & 1.00 & 0.16 & \maxColor{ $\boldsymbol{0.48}$ } \\  \cline{2-8}
      & recall                                                  &  1.00 & 0.01 & \maxColor{ $\boldsymbol{0.11}$ } & 1.00 & 0.02 & \maxColor{ $\boldsymbol{0.06}$ } \\ \hline
   \end{tabular}
   \label{tab:refind}
\end{table}
\normalsize

\begin{figure*}
    \setlength{\tabcolsep}{0.8pt}
    \hspace{0.75em}
    \subfloat[The page-level avg. story replacement rates for \textit{General} SERP collections show a direct relationship between page number and story replacement rate - the higher the page number, the higher the story replacement rate, and vice versa.]{{ \includegraphics[width=0.47\textwidth]{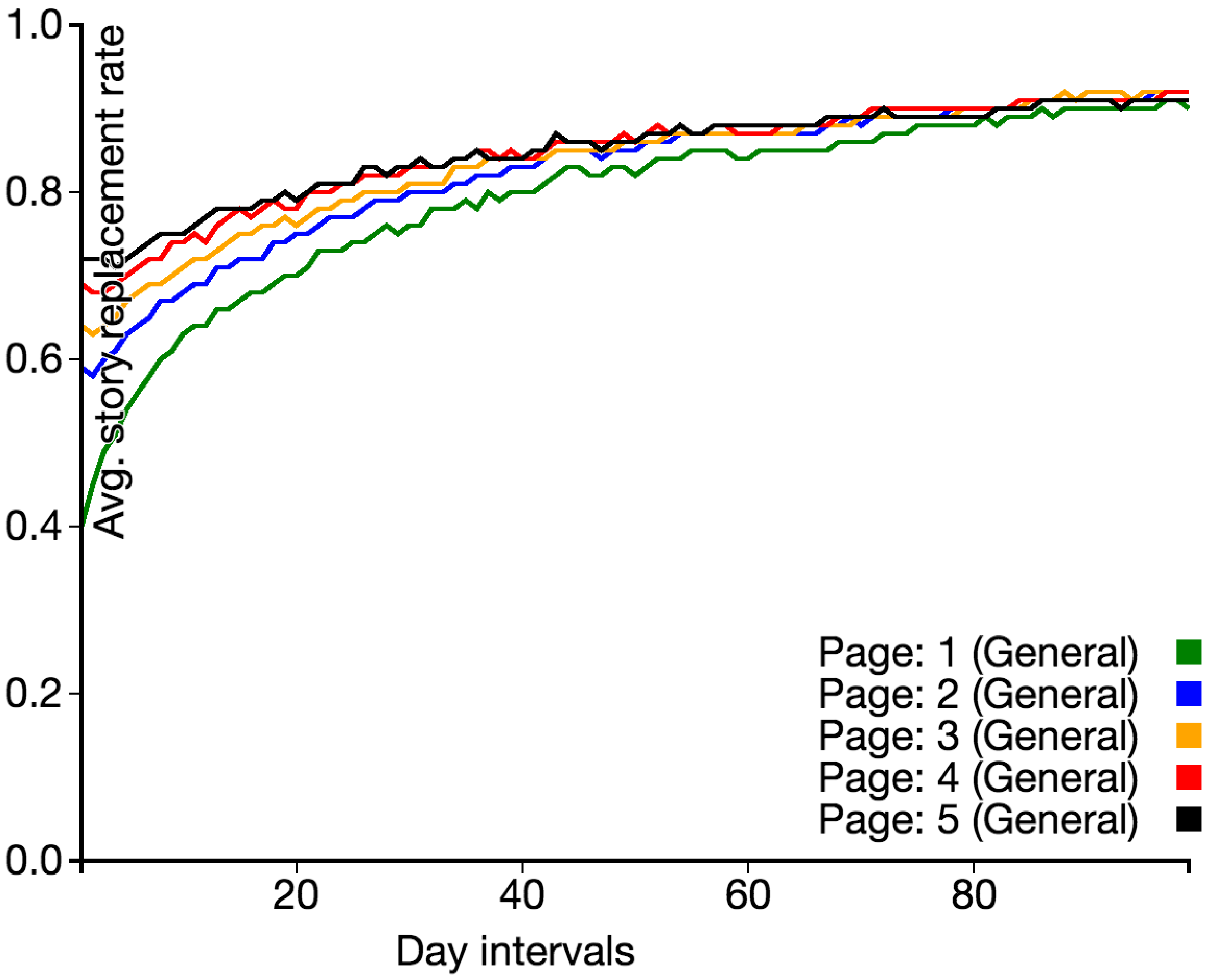} }}
    \hspace{0.75em}
    \subfloat[The page-level avg. story replacement rates for \textit{News} vertical SERP collections show an inverse relationship between page number and story replacement rate - the higher the page number, the lower the story replacement rate, and vice versa.]{{ \includegraphics[width=0.47\textwidth]{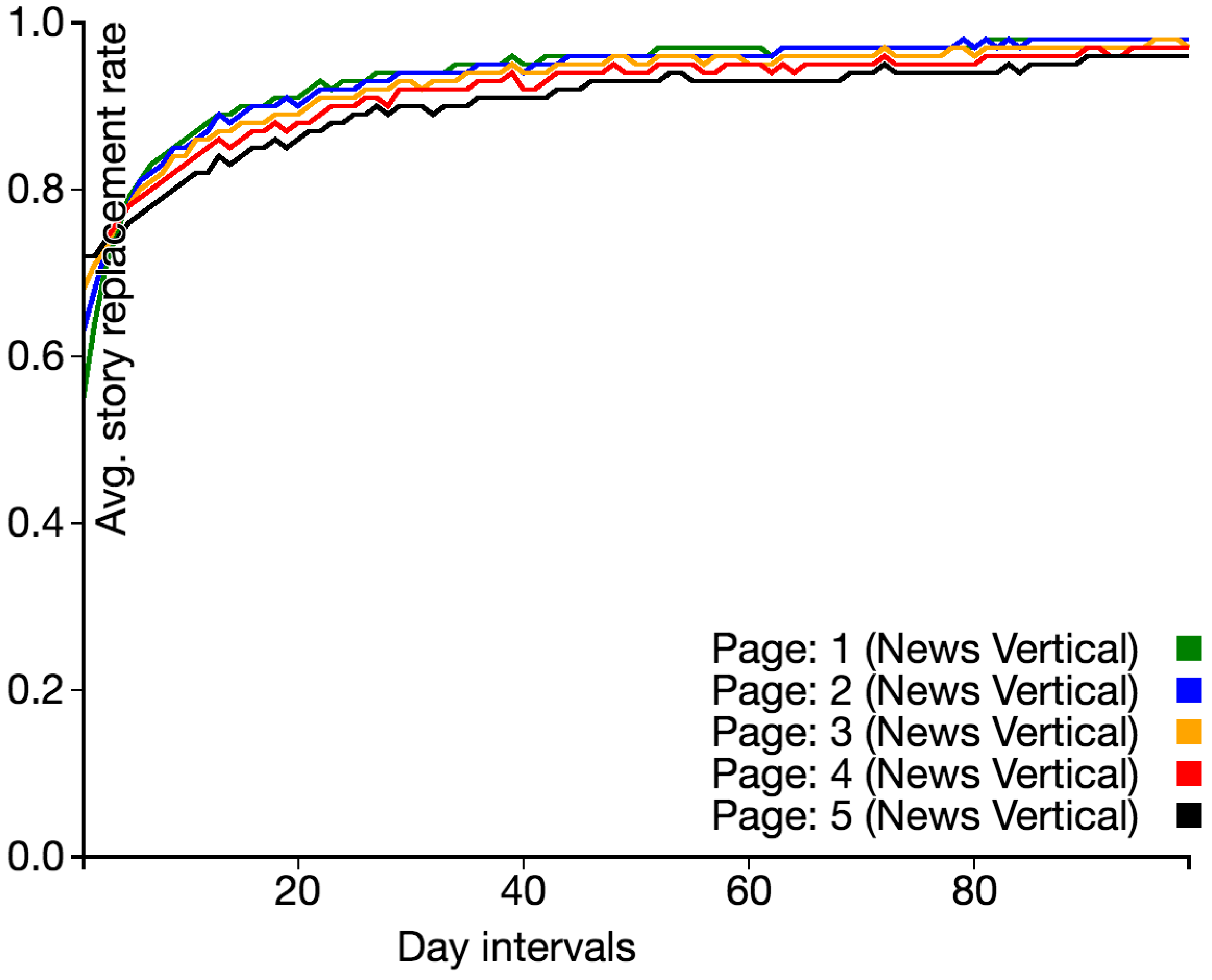} }}
    \hspace{0.75em}
    \subfloat[Similar to the page-level avg. story replacement rate, the page-level avg. new story rate for \textit{General} SERP collections show a direct relationship between page number and new story rate.]{{ \includegraphics[width=0.47\textwidth]{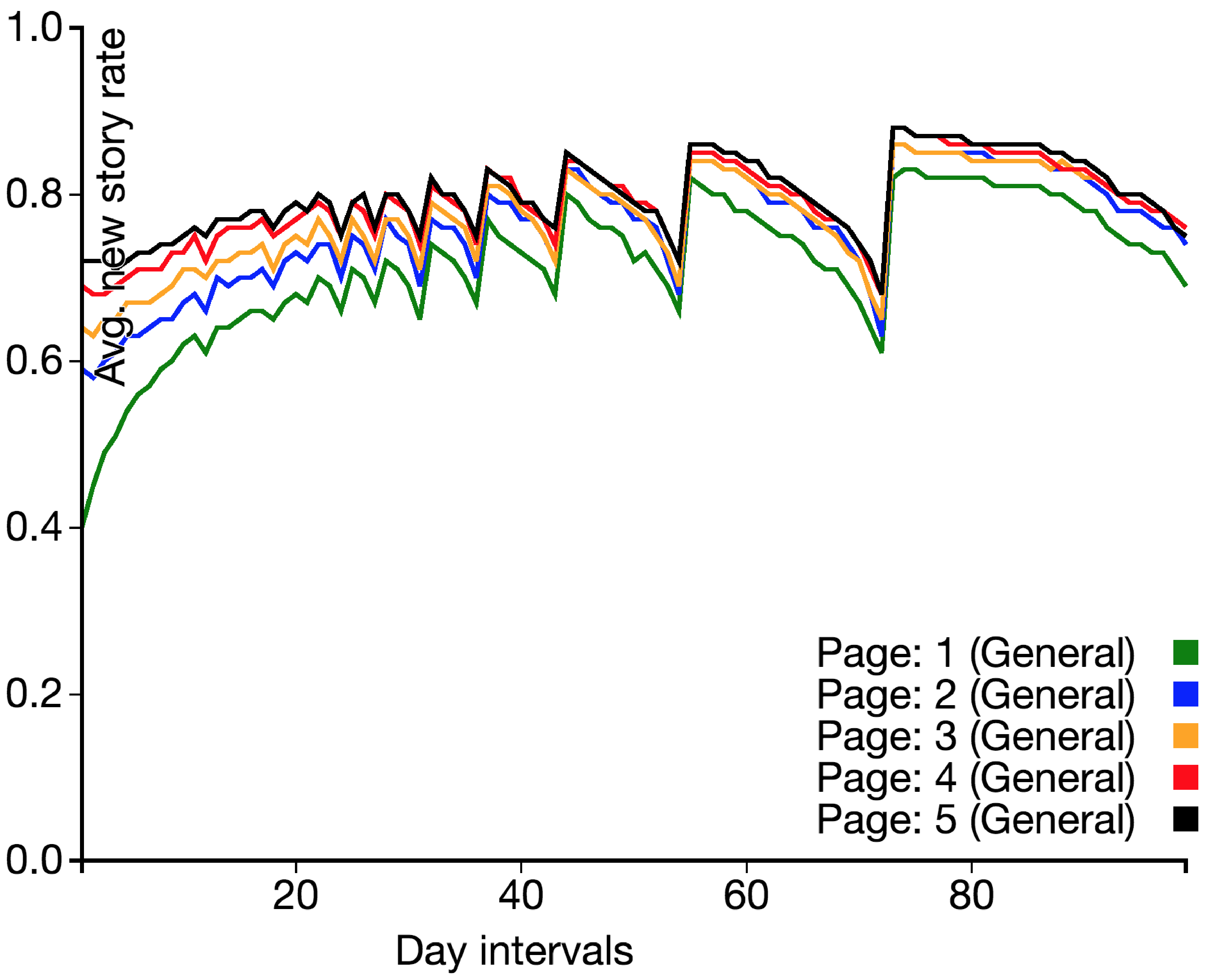} }}
    \hspace{0.75em}
    \subfloat[Similar to the page-level avg. story replacement rate, the page-level avg. new story rate for \textit{News} vertical SERP collections show an inverse relationship between page number and new story rate.]{{ \includegraphics[width=0.47\textwidth]{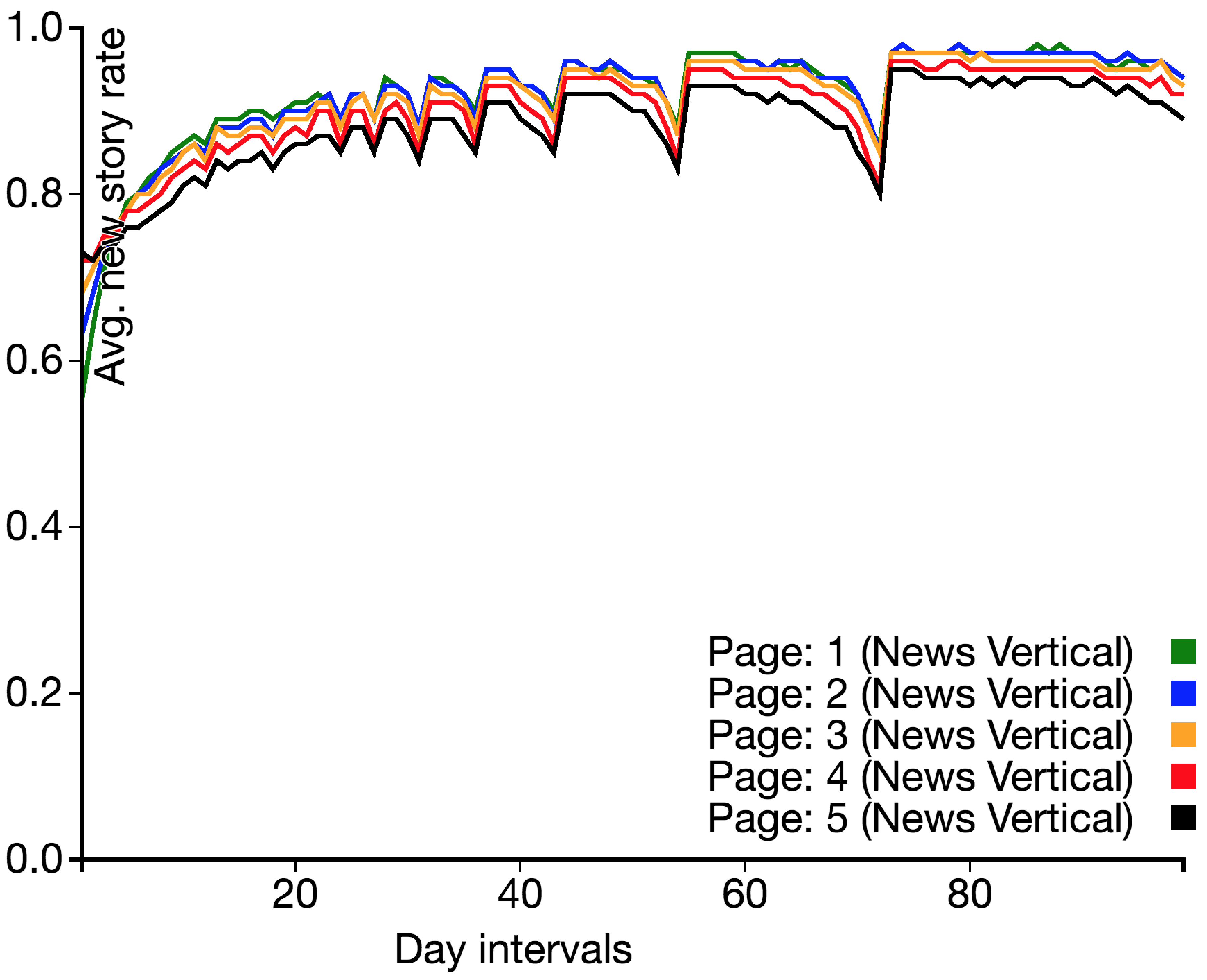} }}
    \caption{a \& b: Page-level new story rates for \textit{General} and \textit{News} vertical SERPs. c \& d: Page-level story replacement rates for \textit{General} and \textit{News} vertical SERPs.}%
    \label{fig:pageLevelStats}%
\end{figure*}
\begin{figure*}
    \setlength{\tabcolsep}{0.8pt}
    \subfloat[Prob. of finding a story after variable number of days on pages (1-5) for \textit{General} SERP shows direct relationship between page number and prob.]{{ \includegraphics[width=0.47\textwidth]{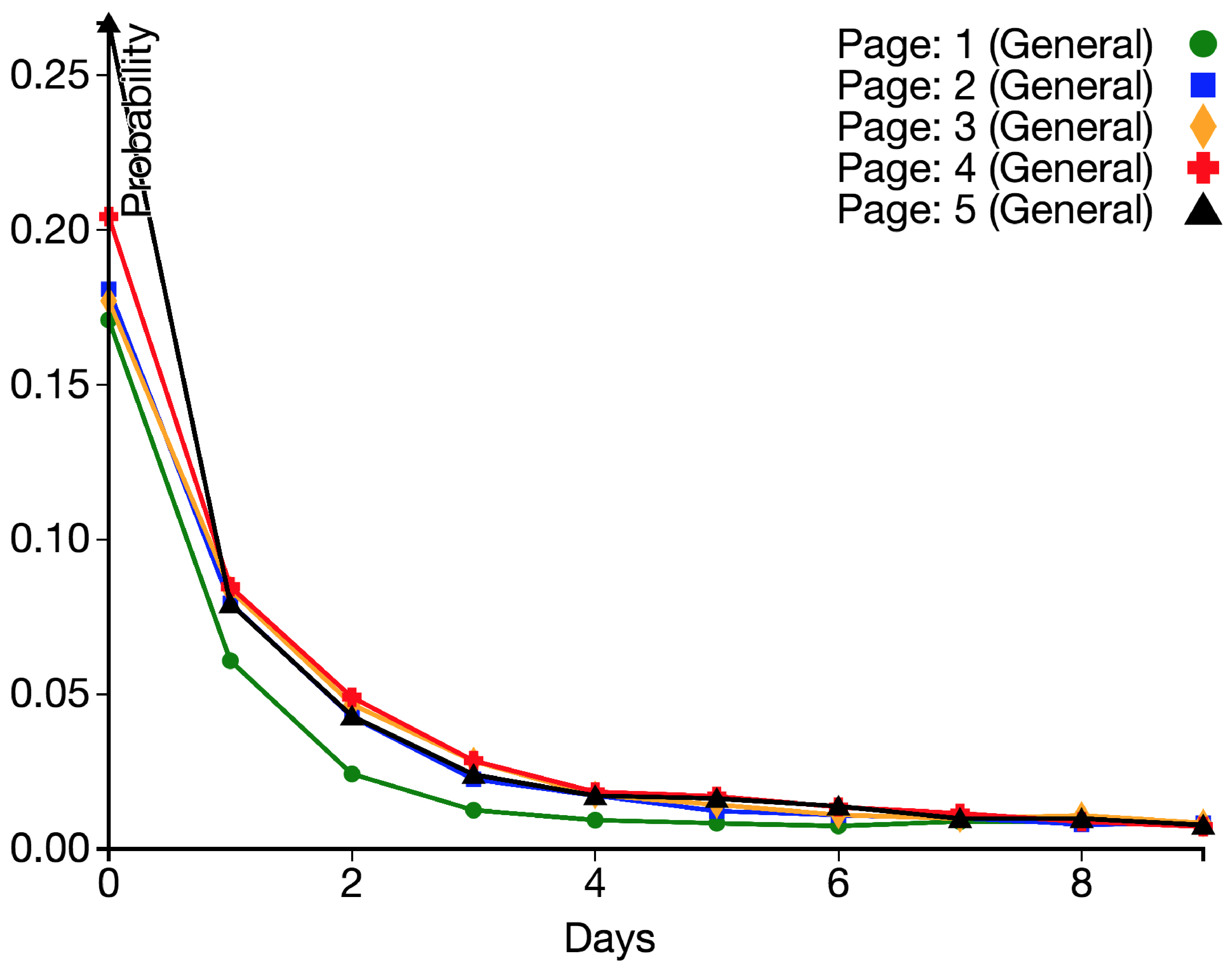} }}
    \qquad
    \subfloat[Prob. of finding a story after variable number of days on pages (1-5) for \textit{General} SERP shows inverse relationship between page number and prob.]{{ \includegraphics[width=0.47\textwidth]{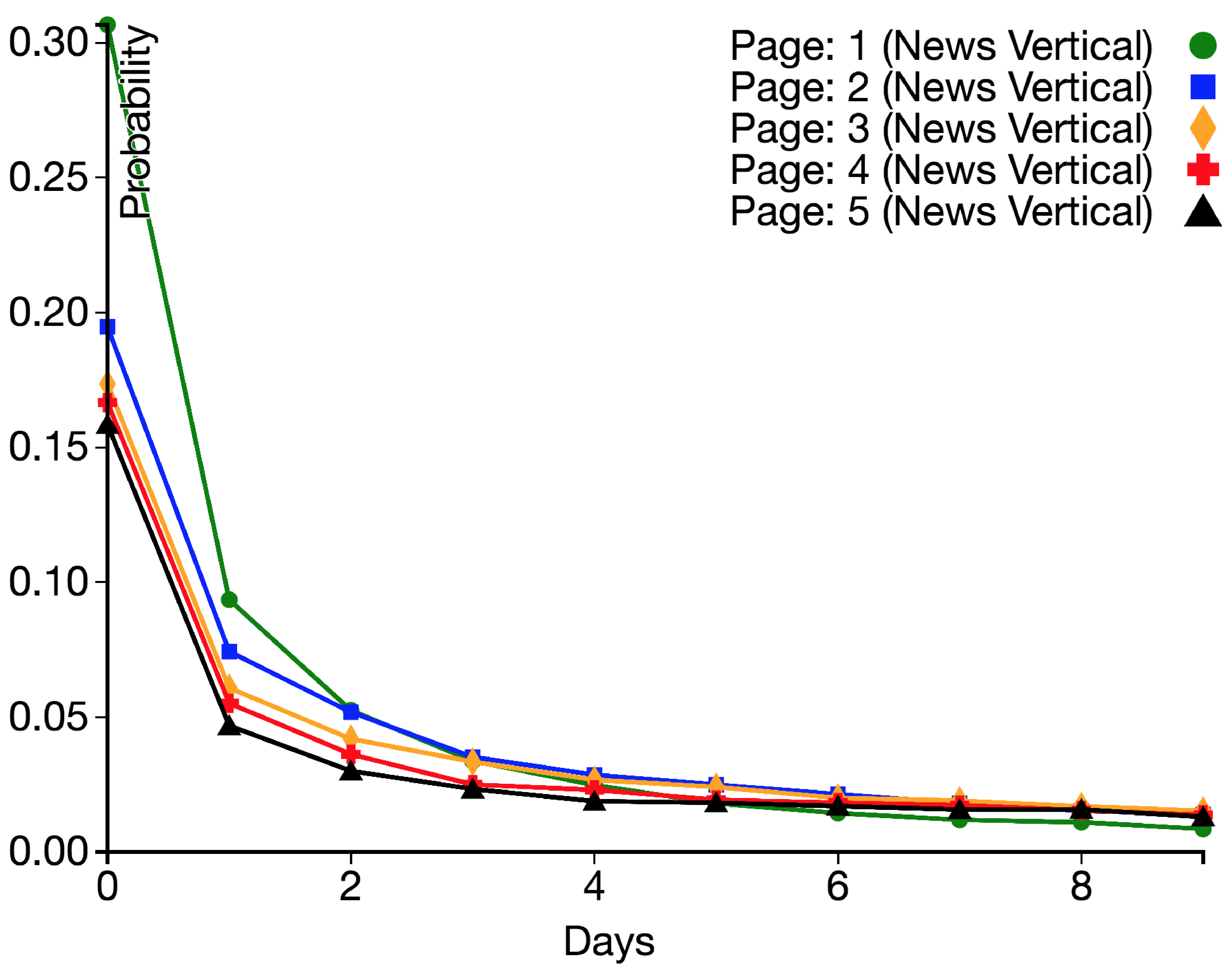} }}
    \caption{a \& b: Page-level probability of finding the URI of a story over time.}%
    \label{fig:pgLevelProb}%
\end{figure*}
\begin{figure}
   \centering
   \includegraphics[width=0.47\textwidth]{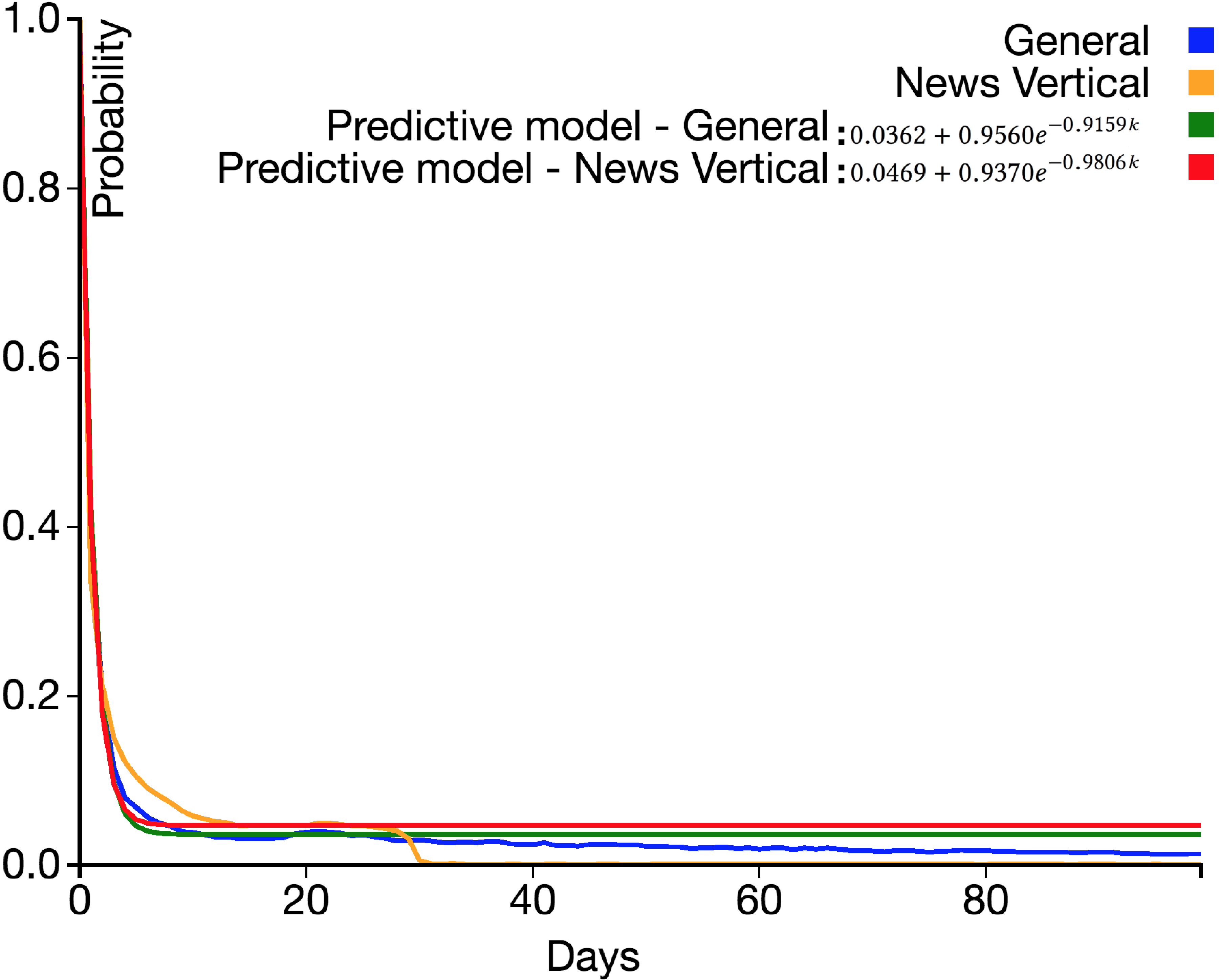}
   \caption{Prob. of finding an arbitrary story for \textit{General} and \textit{News} vertical SERPs was modeled with two best-fit exponential functions. In general, the probability of finding the URI of a news story on the \textit{General} SERP is higher (lower new story rate) than the probability of finding the same URI on the \textit{News} vertical SERP (due to its higher new story rate).}
   \label{fig:bestFit}%
\end{figure}

Here we present the results for each of the respective measures introduced in Subsection \ref{sec:primitiveMeasures}.
\subsection{\textbf{Story replacement rate, new story rate, and page level new story rate}}
Table \ref{tab:avgStoryReplacementRate} and \ref{tab:avgNewStoryRate} show the average story replacment rate and new story rate, respectively over time (daily, weekly, and monthly) for both the \textit{General} and \textit{News} vertical SERPs. For both \textit{General} and \textit{News} vertical SERPs, we can see that the average story replacement rate was similar to the new story rate, and both increased with time. They also show that the story replacement and new story rates are strongly dependent on the topic. For example, the \textit{Hurricane Harvey} natural disaster showed a lower daily average story replacement rate (0.21) and new story rate (0.21) compared to the \textit{Trump-Russia} event. This event maintained the highest daily (0.54), weekly (0.79), and monthly (0.92) average story replacement and new story rates (0.54 - daily, 0.78 - weekly, and 0.83 - monthly). Unlike natural disasters which have a well-defined timeframe, this on-going political event does not have a well-defined timeframe and has undergone multiple event cycles - from the firing of the FBI Director James Comey in May 2017 to the indictment of former Trump Campaign Chair Paul Manafort in October 2017. Similar to the \textit{General} SERP, the average story replacement rate and new story rate for the \textit{News} vertical SERP increased with time but at much faster rates. These results show us that the timing of collection building efforts that utilize SERPs is critical especially for rapidly evolving events with undefined timeframes. Since these events produce newer stories continuously, collection building must be continuous in order to capture the various cycles of the event.

Figs. \ref{fig:pageLevelStats}a \& c show that the average story replacement rate and average new story rate differed across various pages for the \textit{General} SERP. There was a direct relationship between page number and story replacement rate (or new story rate) - the higher the page number, the higher the story replacement rate (or new story rate), and vice versa. The direct relationship may be due to fact that higher order pages (e.g., pages 4 and 5) are more likely to receive documents from lower order pages (e.g, page 1 -- 3) than the opposite. For example, the probability of going from page 1 to page 5 was 0.0239 while the probability of going from page 5 to page 1 was 0.0048. The lower order pages have the highest quality on the SERP, thus, there is high competition within documents to retain their position on a lower order page (high rank). The competition in the higher order pages is less, therefore, when documents from the lower order pages lose some rank, they may fall into the higher order pages thereby increasing the new story rate of higher order pages. The \textit{News} vertical SERP showed an inverse relationship between the page number and the story replacement rate (or new story rate) (Fig. \ref{fig:pageLevelStats}b \& d) even though the probability of going from a page 1 to page 5 (0.0801) was more likely than the opposite (0.0009). This may be due to some unseen mechanism in the \textit{News} vertical SERP.

\subsection{\textbf{Probability of finding a story}}
Table \ref{tab:probFindingStory} shows the probability of finding the same story after one day, one week, and one month (from first observation) for \textit{General} and \textit{News} vertical SERP collections. The probability of finding the same URI of a news story with the same query decreased with time for both SERP collections. For the \textit{General} SERP, the probability of the event that a given URI for a news story is observed on the SERP when the same query is issued one day after it was first observed ranged from 0.34 -- 0.44. When the query was issued one week after, the probability dropped to from 0.01 -- 0.11, one month after - 0.01 -- 0.08. The probability of finding the same story with time is related to the rate of new stories: for a given time interval, the higher the rate of new stories, the lower the chance of observing the same story, because it is more likely to be replaced by another story. For example, compared to the \textit{manchester bombing} collection, the \textit{hurricane irma} collection produced a lower (0.34) probability (vs. \textit{manchester bombing} - 0.44) of finding the same story after one day due to its higher (0.79) new story rate after one day (vs. \textit{manchester bombing} - 0.52). The probability of observing the same news story on the \textit{News} vertical SERP declined with time, but at a much faster rate compared to the \textit{General} SERP. In fact, Table \ref{tab:probFindingStory} shows that for all seven topics in the dataset, the probability of finding the same story on the \textit{News} vertical when the query was re-issued one month after was marginal (approximately 0). This is partly because the \textit{News} vertical SERP collections produced higher story replacement and new story rates than the \textit{General} SERP collections.

In order to generalize the probability of finding an arbitrary URI as a function of time (days), we fitted a curve (Fig. \ref{fig:bestFit}) over the union of occurrence of the URIs in our dataset with an exponential model. The probability $P_{s, sp}(k)$ of finding an arbitrary URI of a news story $s$ on a SERP $sp \in \{General, News Vertical\}$, after $k$ days is predicted as follows:
\begin{eqnarray*}
  P_{s, General}(k) = 0.0362 + 0.9560e^{-0.9159k}\\
  P_{s, NewsVertical}(k) = 0.0469 + 0.9370e^{-0.9806k}
  \label{eqn:Gpredict}
\end{eqnarray*}Also, similar to the story replacement and new story rates, for the \textit{General} SERP, the results showed a direct relationship with the page number and probability of finding news stories over time (Fig. \ref{fig:pgLevelProb}a). For the \textit{General} SERP, higher order page numbers (e.g., 4 and 5) produced higher probabilities of finding the same stories compared to lower order (e.g., 1 and 2) pages. This might be because during the lifetime of a story, the probability of the story going from a lower order (high rank) page to a higher (low rank) order page is higher than the opposite - going from higher order page to lower order page (climbing in rank). For example, the probability of going from page 1 to page 5 was higher (0.0239) than the probability of going from page 5 to page 1 (0.0048). However, collections from \textit{News} verticals showed that the lower the page number, the higher the probability of finding news stories (inverse relationship) even though the probability of falling in rank (lower order page to higher order page) is higher than the probability of climbing in rank (higher order page to lower order page).

\subsection{\textbf{Distribution of stories over time across pages}}
\label{res:disOfStoriesPerTime}
Fig. \ref{tab:tempDist} shows how the temporal distributions differs typically between \textit{General} and \textit{News} vertical SERP collections. There are two dimensions in the figure: days (x-axis) and URIs of stories (y-axis). A single dot in the figure indicates that a specific story occurred at that point. The temporal distritution is a reflection of the new story rate, but at a granular (individual) story level. \textit{General} SERP collections had lower new story rates, thus produced stories with a longer longer lifespan than \textit{News} vertical SERP collections. In Fig. \ref{tab:tempDist}, this is represented by a long trail of dots. Since \textit{News} vertical collections had higher story replacement and new story rates, they produced documents with shorter lifespans. For example, Fig. \ref{tab:tempDist}a contrasts the denser (longer lifespan) temporal distribution of the ``hurricane harvey'' \textit{General} SERP collection to the sparser ``trump russia'' \textit{General} SERP collection (Fig. \ref{tab:tempDist}c). The ``trump russia'' collection  produced new documents on average at a rate of 0.54 (daily) to 0.83 (monthly), compared to the ``hurricane harvey'' collection (daily - 0.21, and monthly - 0.51). Similarly, since documents from the ``trump russia'' collections were rapidly replaced (story replacement rate: 0.54 -- 0.92) with newer documents, they mostly did not persist on the SERP.

Figs. \ref{tab:normTempDistG} and \ref{tab:normTempDistNV} show how URIs moved across pages over time. The rows represent the URIs and the columns represent the pages in which the URIs were observed on a specific day. A single cell represents the page in which a URI occured on a specific day. For example, the first cell (row 0, column 0) of Fig. \ref{tab:normTempDistG} is 1. This means the URI at row 0 was first observed on page 1. Some of the same URIs persist over time within the same page. For example Fig. \ref{tab:normTempDistG}, row 0, shows that the highly ranked Wikipedia page\footnote{https://en.wikipedia.org/wiki/Manchester\_Arena\_bombing} of the \textit{Manchester bombing} event was seen for 24 consecutive days on the first page of the SERP, was not seen (within page 1-5) on the 25th day, and then seen for 13 consecutive days (still on page 1). Fig. \ref{tab:normTempDistG} also shows the increase/decrease in ranks for stories. For example, in Fig. \ref{tab:normTempDistG}, row 4, the URI\footnote{http://www.dailymail.co.uk/news/article-4578566/Evidence-Nissan-linked-Manchester-bombing.html} was first observed on page 5, the next day it increased in rank to page 1, skipping (2-4). The page-level temporal distribution also shows that some stories go directly from page 5 to 1. In contrast with \textit{General} SERP collections, the temporal distrbution of \textit{News} vertical collections is shorter (Fig. \ref{tab:normTempDistNV}) and reflect the higher story replacement and new story rates of \textit{News} vertical collections.

\subsection{\textbf{Overlap and recall}}
Table \ref{tab:refind} shows that setting the Google date range parameter improves finding stories with respect to the set date range for both \textit{General} and \textit{News} vertical collection. For example, for the ``healthcare bill'' \textit{General} SERP collection, the \textit{Jan-2018} collection which was created (2018-01-11) by making a default search (without) setting the date range had an overlap rate of 0.06 with respect to the collection of documents created in June 2017 (\textit{June-2017}). In contrast, the collection created the same day (2018-01-11) by setting the date range parameter to June 2017 (2017-06-01 to 2017-06-30) had a much higher overlap rate of 0.60. This is the case across all collection topics, especially for topics with lower new story rates (0.27 - 0.46) such as ``manchester bombing'' (0.82 overlap rate). The \textit{News} vertical collections had lower overlap rates compared to the \textit{General} SERP collections since \textit{News} vertical collection have higher story replacement and new story rates.

Irrespective of the increase in refinding (overlap) new stories that occurs when the date range parameter is set, the recall is poor. Since the SERP only produces a fixed number of documents per page, we only get a small fraction of the documents relevant to the specified date range. The ``healthcare bill'' \textit{June-2017} \textit{General} SERP collection contains 460 documents published in June 2017, collected by extracting URIs from the first five pages of the SERP. A query (``healthcare bill'') issued to the SERP in January 2018, with the date range parameter set to June 2017 increased overlap (refinding stories), but did not increase the number of results - we could only extract at most approximately 50 URIs (first five page). Consequently, across all topics in Table \ref{tab:refind}, both \textit{Jan-2018} and \textit{Jan-2018-Restricted-to-June} collections had recall of under 0.10 except for the ``london terrorism'' topic (maximum recall 0.20). This reaffirms the that idea that collection building or seed selection processes that rely on the SERP must start early and persist in order to maximize recall. To further aid selection of seeds, a simple set of heuristics could identify most of the likely stable URIs (e.g., \textit{wikipedia.org}, \textit{nasa.gov}, \textit{whitehouse.gov}) as well as URIs likely to quickly disappear from the top-k SERPs (e.g., \textit{cnn.com} or \textit{nytimes.com}, followed by a long path in the URI). The archivist could give priority to the latter URIs, knowing that the former URIs will continue to be discoverable via Google.
\section{Future work and Conclusions}
Our findings motivate the need for instant and persistent collection building. The next stage of this research is the implementation of a collection building system that extracts URIs from SERPs. The system may be triggered semi-automatically by a curator for an important event or automatically by an event detection system. An event detection system could listen to traffic in news and social media for clusters of interest, identify the events of interest, and initiate a collection building process from SERPs. In addition to the implementation of such a collection building system, it is important to investigate the kinds of collections, topics or events most suitable for SERPs. For example, this research focused on news collections, but further research is required to assess the effectiveness of using SERPs for other kinds of collections.

Collection building offers a way of preserving the historic record of important events. This involves collecting URIs of web pages that are relevant to a predefined set of topics. It is crucial for collections to capture all the stages (oldest to newest) of events and not only the recent stories. Search engines provide an opportunity to build collections or extract seeds, but tend to provide the most recent documents. As a first step toward a larger effort of generating collections from SERPs, we sought to gain insight on the dynamics of refinding stories on SERPs. Our findings illustrate the difficulty in refinding news stories as time progresses: on average, the daily rate at which stories were replaced on the Google \textit{General} SERP ranged from 0.21 -- 0.54, weekly - 0.39 -- 0.79, and monthly - 0.59 -- 0.92. The Google \textit{News} vertical SERP showed even higher story replacement rates, with a daily range of 0.31 -- 0.57, weekly - 0.54 -- 0.82, and monthly - 0.76 -- 0.92. We also showed that the probability of finding the same news story diminishes with time and is query dependent. The probability of finding the same news story with the same query again, one day after the first time the story was first seen ranged from 0.34 -- 0.44. If one waited a week, or a month and issued the same query again, the probability of finding the same news story drops to 0.01 -- 0.11. The probability declines even further if we used the \textit{News} vertical SERP due to its higher story replacement and new story rates. The probability for finding the URI of a news story was further generalized through our provision of two predictive models that estimate this probability as a function of time (days). Discoverability may be improved by instructing the search engine to return documents published within a temporal range, but this information is not readily available for many events, and we discover only a small fraction of relevant documents since the count of search results are restricted. These findings collectively express the difficulty in refinding news stories with time, thus motivates the need for collection building processes that utilize the SERP to begin early and persist in order to capture the start and evolution of an event. Our research dataset comprising of 151,602 (33,432 unique) links extracted from the Google SERPs for over seven months, as well as the source code for the application utilized to semi-automatically generate the collections are publicly available \cite{jcdl2018Repo}.
\section*{Acknowledgements}
This work was made possible by IMLS LG-71-15-0077-15, and help from Christie Moffat at the National Library of Medicine.
\clearpage
\bibliographystyle{ACM-Reference-Format}
\balance
\bibliography{NwalaJCDL2018} 

\end{document}